\RequirePackage[T1]{fontenc}
\documentclass[12pt]{article}

\usepackage[height=8.85in,width=6.25in]{geometry}

\usepackage[utf8]{inputenc}
\usepackage{amsmath}
\usepackage{amssymb}
\usepackage{mathtools}
\numberwithin{equation}{section}
\usepackage{slashed}
\usepackage{braket}
\usepackage[svgnames,dvipsnames]{xcolor}
\usepackage[colorlinks,citecolor=DarkGreen,linkcolor=FireBrick,urlcolor=FireBrick,linktocpage,breaklinks=true]{hyperref}
\usepackage{cite}
\usepackage{graphicx}
\usepackage{tikz}

\usepackage{times}
\usepackage{courier}
\usepackage{bm}
\usepackage{subfig}
\usepackage{here}
\usepackage{stmaryrd}

\definecolor{identifiercolor}{rgb}{.4,.6,.56}
\definecolor{stringcolor}{gray}{0.5}
\definecolor{inactivecolor}{rgb}{0.15,0.15,0.5}

\usepackage{listings}
\newcommand{\code}[1]{\lstinline{#1}}

\def\Co{\mathrm{Co}}
\def\Suz{\mathrm{Suz}}
\def\bC{\mathbb{C}}
\def\bR{\mathbb{R}}
\def\bZ{\mathbb{Z}}
\def\cH{\mathcal{H}}

\def\sS{\mathsf{S}}
\def\sT{\mathsf{T}}
\def\cO{\mathcal{O}}
\def\cZ{\mathcal{Z}}
\def\Nequals#1{$\mathcal{N}{=}#1$}

\def\Arf{\mathop{\mathrm{Arf}}}
\DeclareMathOperator{\Hom}{Hom}
\DeclareMathOperator{\K}{K}
\DeclareMathOperator{\KO}{KO}
\DeclareMathOperator{\TMF}{TMF}
\DeclareMathOperator{\tr}{tr}

\def\NS{\mathrm{NS}}
\def\rR{\mathrm{R}}
\def\rX{\mathrm{X}}
\def\rY{\mathrm{Y}}
\def\iu{i}

\def\mug{\bm{\lambda}}
\def\muh{\bm{\mu}}

\begin{document}

\begin{titlepage}

\begin{flushright}

\end{flushright}

\vskip 3cm

\begin{center}

{\Large \bfseries The Art of Orbifolding\\[1em]
by Huge Non-Abelian Groups}

\vskip 1cm
Jan Albert$^1$, Masaki Okada and Yuji Tachikawa$^2$
\vskip 1cm

\begin{tabular}{ll}
$^1$ & Princeton Center for Theoretical Science, \\
& Princeton University, Princeton, NJ 08544, USA\\
$^2$ & Kavli Institute for the Physics and Mathematics of the Universe (WPI), \\
& University of Tokyo,  Kashiwa, Chiba 277-8583, Japan
\end{tabular}

\vskip 2cm

\end{center}

\noindent 

We compute torus partition functions of orbifolds
of two-dimensional chiral fermions by huge non-Abelian groups involving
the Conway group, one of the sporadic finite simple groups.
This seemingly standard problem turns out to contain,
somewhat unexpectedly,
multiple major issues hitherto unresolved, 
both at the level of the principle
and at the level of practice.
We describe in detail how they can be overcome.

Although the methods described in this paper are generally applicable,
the choice of the particular orbifolds we consider in this paper
was motivated by the search for the periodicity element of topological modular forms.
We will provide a brief introduction to this question,
to explain why all the techniques in this paper had to be developed.
Unfortunately, the orbifolds we considered turned out \emph{not} to be the periodicity element.

\end{titlepage}

\setcounter{tocdepth}{2}
\tableofcontents

\section{Introduction and summary}
In this paper we compute the torus partition functions of orbifolds of two-dimensional chiral fermions by huge non-Abelian groups.
As a concrete example, we first consider the $\bZ_2$ orbifold 
$V=F_{24}/\bZ_2$,
where $F_n$ is the theory of $n$ free Majorana-Weyl fermions.
This theory $V$ is known to have supersymmetry
with a commuting action of the Conway group $\Co_0$,
which is a central extension by $\bZ_2$ of $\Co_1$, one of the 26 sporadic finite simple groups.
We then consider the orbifold \begin{equation}
V^{24}/[S_{24}\times \Co_0],
\label{eq:our-orbifold}
\end{equation}
where 
$S_{24}$ permutes 24 copies of the theory $V$.
In the end, this is simply an orbifold
of $F_{24\times 24}=F_{576}$ by 
a non-Abelian group $(\bZ_2^{24}\rtimes S_{24})\rtimes \Co_0$,
of order $\sim 8.66\times 10^{49}$.
We compute the partition function of this orbifold on the torus 
with periodic spin structure, and find that it is a constant $81270$.

The main motivation to consider this particular orbifold came from our search for the periodicity element of topological modular forms,
but the methods we develop for this purpose are generally applicable.
Therefore we start with two introductions which can be read independently,
depending on the interest of the reader.
We first describe in Sec.~\ref{intro:orbifold} why the seemingly simple problem of computing the torus partition function of orbifolds of free fermions contains a number of major issues to be overcome.
We then explain in Sec.~\ref{intro:tmf} why we were led to consider this 
particular orbifold and its variants treated in this paper, 
from a question concerning the relation between the mathematical theory of topological modular forms and the physics of two-dimensional conformal field theories.

\subsection{Issues in computing orbifold partition functions}
\label{intro:orbifold}

Given a theory $T$ with a finite symmetry group $G$, the torus partition function of the orbifold $T/G$ is given by
\begin{equation}
Z[T/G](\tau) = \frac{1}{|G|}\sum_{gh=hg} Z[T]^h_g(\tau),
\label{eq:naive}
\end{equation}
where $Z[T]^h_g(\tau)$ is the partition function of $T$ 
on a torus with complex structure $\tau$ with
holo\-nomies $g$ along the spatial cycle and $h$ along the temporal cycle.
(Here and in the following, we also call the spatial cycle the horizontal cycle, and the temporal cycle the vertical cycle.)
Concretely, we take $T$ to be the theory $F_{2N}$ of  $2N$ free Majorana-Weyl fermions.
It has a symmetry $SO(2N)_1$, and we take $G$ to be a non-anomalous finite subgroup of $SO(2N)$.
The two holonomies $g,h\in SO(2N)$ commute. 
Assuming that $g,h$ are simultaneously diagonalizable, 
we can go to a basis of $\bR^{2N}\simeq \bC^{N}$ where $g$, $h$
act diagonally with eigenvalues $e^{2\pi i \lambda_a}$, $e^{2\pi i\mu_a}$  with absolute value 1.
Then the partition function $Z[T]^h_g(\tau)$ is given by a product of $N$ theta functions, \begin{equation}
Z[T]^h_g(\tau) \sim \prod_{a=1}^N \theta^{\mu_a}_{\lambda_a}(\tau).
\end{equation}
The reader should not worry that our notation is not yet precise;
the point here is to point out where lies the problem.

Would this make the expression \eqref{eq:naive} concrete enough to be computed?
Unfortunately not. 
There are two main issues, one practical issue for non-Abelian groups,
and one issue in principle even for Abelian groups.

The practical issue is to enumerate the commuting pairs $(g,h)$ in $G$.
For a non-anomalous symmetry, two commuting pairs $(g,h)$ and $(g',h')$ related 
by conjugation by an element $k\in G$, i.e.~those satisfying $(g',h')=(kgk^{-1},khk^{-1})$, give the same contribution to \eqref{eq:naive}.
Therefore what we need is to enumerate the conjugacy classes of commuting pairs $(g,h)$ in $G$.
Conjugacy classes of elements in $G$ are often well-studied and can readily be found in the literature, such as \cite{ATLAS}, 
but conjugacy classes of commuting pairs are much less studied. 
We need to do this by ourselves using computer algebra systems,
and for our particular choice of $G$, it turns out to require 
the use of a computer cluster.

The second issue is more of a principle, and exists even for Abelian groups.
The theory $F_{2N}$ of $2N$ free Majorana-Weyl fermions has an $SO(2N)_1$ symmetry.
But the fact that it is of level 1 means that the symmetry is anomalous.
This manifests itself in the fact that the factors of theta functions $\theta^{\mu_a}_{\lambda_a}(\tau)$ in $Z[T]^h_g(\tau)$
are not invariant under the modular transformation on $\tau$
or under the shift of the angles $\mu_a$ and $\lambda_a$ by integers.
They instead transform with nontrivial phases, which are well-defined once we fix all the conventions.
This means that by simply pulling back the $SO(2N)_1$ anomaly to $G$, 
we will always have a nontrivial anomaly cocycle for the $G$ symmetry.
What is special when $G$ is non-anomalous is that the anomaly cocycle
is a coboundary.
Then, we can compensate the anomalous phases of the products of theta functions by a suitable choice of phases $\alpha(g,h)$ for each commuting pair $(g,h)$,
so that we have \begin{equation}
Z[T]^h_g(\tau) = \alpha(g,h) \prod_{a=1}^N \theta^{\mu_a}_{\lambda_a}(\tau).
\end{equation}
Once the phases $\alpha(g,h)$ are chosen, the expression \eqref{eq:naive} is well-defined and can be computed.

The main problem here is that there is no known general method to obtain the phases $\alpha(g,h)$.\footnote{
The Abelian cases should have been implicitly worked out in the literature 
when Abelian orbifolds of free fermions were studied in the late 1980s
in the context of free-fermionic constructions of string theory,
although the authors have not tracked down the relevant references.
See \cite{Freed:1987qk} as a closely related topic from this time period.}
The GSO projection used in the superstring theory 
is in some sense the simplest nontrivial case, when $G=\bZ_2$.
In this case, the phases $\alpha(g,h)$ are usually determined by requiring
the modular invariance of the resulting orbifold partition function.
But this method is useful only when any commuting pair $(g,h)$ 
can be mapped to a pair $(e,h')$, where $e$ is the trivial holonomy, by the $SL(2,\bZ)$ transformation.
This is not true in general, even for Abelian groups which are not cyclic.\footnote{%
The condition for any commuting pair $(g,h)$ to be mapped to a pair $(e,h')$ by 
some $SL(2,\bZ)$ transformation is that any Abelian subgroup of $G$ is cyclic.
It is known that this condition is equal to the condition that $H^*(BG;\bZ)$ is periodic, i.e.~there exists a class $u\in H^k(BG;\bZ)$ with $k>0$ such that the cup product with $u$ gives an isomorphism $H^n(BG;\bZ)\to H^{n+k}(BG;\bZ)$ for $n\ge 1$.
Furthermore, such finite groups have been completely classified by Suzuki and Zassenhaus. For a textbook account, see \cite[Chapter IV.6]{AdemMilgram}.
If $G$ acts on a sphere $S^n$ freely, then $G$ is periodic.
For example, finite subgroups of $SU(2)$ are periodic.
But the converse is not true, and there are periodic groups which do not act freely on any sphere.
}

In the last several years, our understanding of the anomaly of finite groups,
both in bosonic and fermionic systems, has been greatly advanced.
In the bosonic case the anomaly is classified by the group cohomology $H^3(BG;U(1))$, which can be expressed in terms of cochains.
Here we deal with free fermions, and therefore need anomalies in the fermionic case.
Then it is classified by the Pontryagin dual of the spin bordism group $\Hom(\Omega^\text{spin}_3(BG),U(1))$,
whose explicit cochain-level description is also known.
(In the following, we call these cochain-level descriptions simply as cochains,
by a slight abuse of terminology.)
As the anomaly of $SO(2N)_1$ symmetry of the fermions is also known,
it should in principle be possible to obtain the anomaly cochains
of the subgroup $G$ of $SO(2N)$, and trivialize it when it is a coboundary.
But, so far, only the anomaly cochains of Abelian subgroups of $SO(2N)$
have been worked out in the literature \cite{Okada:2025kie}.
Furthermore, even if such an anomaly cochain could be obtained for our non-Abelian group $G$, 
it is unclear how to trivialize it in practice, as the cochains are defined on a huge set.

We will resolve this issue more indirectly,
by imposing subtler consistency conditions 
than the modular invariance of the orbifold partition function.
First, we partially settle the determination of the phases $\alpha(g,h)$
when $G$ is of the form $\bZ_n \times \bZ_m$.
Applying this to the subgroup generated by $g$ and $h$ of $G$,
we can pin down $\alpha(g,h)$ for each $(g,h)$ to a finite set of possibilities.
Second, for each $g$-twisted sector of the original theory $T$,
we demand that the action of the centralizer $C_G(g)$ of $g$ in $G$ 
is well-defined. 
This puts further constraints on the phases $\alpha(g,h)$.
Finally, we require that $\alpha(g,h)$ be universal across
various theories $T$ with the same symmetry group $G\subset SO(2N)_1$,
which is a natural requirement since $\alpha(g,h)$ should in principle
be determined by trivializing the anomaly cocycle of $G$ obtained by 
pulling it back from the anomaly of $SO(2N)_1$.
Luckily for us, these three conditions turn out to be sufficient 
to determine $\alpha(g,h)$ uniquely for our choice of $G$.
We will describe the process in detail in this paper.

\subsection{A central question concerning topological modular forms}
\label{intro:tmf}

Let us now explain our motivation for considering the particular orbifold \eqref{eq:our-orbifold}.
A central concept in algebraic topology is the notion of generalized cohomology theories. 
Two classic examples are complex K-theory $\K^n(X)$ and real K-theory $\KO^n(X)$, which are defined for any topological space $X$.
The former is periodic modulo 2, $\K^{n+2}(X)\simeq \K^n(X)$, 
while the latter is periodic modulo 8, $\KO^{n+8}(X)\simeq \KO^n(X)$.
Not only being important in mathematics, they also appear in many contexts in physics, 
e.g.~in the study of D-brane charges and 
also in the classification of free  topological insulators and superconductors.

Mathematicians have constructed a generalized cohomology theory,
which in some sense arises as a natural next step after the ordinary cohomology and the real K-theory, called topological modular forms (TMF).
In contrast to the complex and real K-theory, both of which have a natural geometric definition in terms of vector bundles,
the only mathematically rigorous definition of TMF is purely algebro-topological, 
and it has been a long-standing question to find a geometric definition of TMF.
A proposal for such a definition was given by 
Segal, Stolz and Teichner \cite{Segal,StolzTeichner1,StolzTeichner2}.
Namely, they proposed that TMF is defined in terms of equivalence classes 
of two-dimensional \Nequals{(0,1)} supersymmetric quantum field theories (SQFTs).
This proposal has been taken up by physicists from around 2018, 
and a number of nontrivial checks and observations have been made.
Slightly more explicitly, elements of $\TMF^n(X)$ are represented by families of SQFTs with gravitational anomaly $n\in \bZ$ parametrized by $X$,
where our convention is to have supersymmetry on the right-movers
and to assign the gravitational anomaly $n=1$
to a single left-moving Majorana-Weyl fermion.
For a recent introductory review on this subject from the physics perspective, 
see the lecture notes by the third author, \cite{TachikawaTMFlecture}.

One of the central questions in this program is the identification of the periodicity element of TMF.
Namely, it is known that TMF is periodic modulo $24^2=576$, $\TMF^{n+576}(X)\simeq \TMF^n(X)$.
This periodicity element should be implemented by  particular SQFTs $X_\pm$
of gravitational anomaly $\pm 576$,
satisfying the condition that $X_+ \otimes X_-$ is equivalent to 
the trivial theory $1$ with single vacuum with zero gravitational anomaly.

It would be most pleasant if the theory $X_-$ is realized by 
a purely right-moving \Nequals{(0,1)} supersymmetric conformal field theory (SCFT) with central charge $c=576/2=288$.
We can flip the orientation and put the same theory 
on the left-movers by forgetting the supersymmetry.
This will then give a theory $X_+:=\overline{X_-}$ 
of gravitational anomaly $+576$.
And we can hope that we might find a suitable deformation 
which makes $\overline{X_-}\otimes X_-$ equivalent to the trivial theory $1$.

A distinguishing feature of such an SCFT $X_-$ is that
its torus partition function with periodic spin structure is 
simply given by the constant $1$, 
without any dependence on the modular parameter $\tau$: \begin{equation}
Z_{X_\pm}(\tau) = 1.
\end{equation}
This is only a necessary condition, but even finding SCFTs satisfying this necessary condition has been a nontrivial task.

One approach to the construction of such SQFTs is the following.
Take the theory $F_{24}$ of $24$ free right-moving Majorana-Weyl fermions, which has gravitational anomaly $-24$ in our convention.
The diagonal $\bZ_2$ symmetry flipping the sign of all fermions is non-anomalous, and we can consider the orbifold \begin{equation}
V:=F_{24}/\bZ_2.
\end{equation}
This theory is known to have a supersymmetry coming from the twisted sector of the $\bZ_2$ orbifold,
and the subgroup of the $SO(24)$ symmetry of $F_{24}$ commuting with this supersymmetry is the Conway group $\Co_0$,
as was found by Duncan \cite{Duncan} and studied further by Duncan and Mack-Crane \cite{DuncanMackCrane}.
Actually, the center $\bZ_2\subset \Co_0$ only acts nontrivially on the R-sector, so the symmetry of the theory $V$ can be taken to be $\Co_1:=\Co_0/\bZ_2$, which acts projectively on the R-sector.
The supersymmetry makes the torus partition function of $V$ with periodic spin structure naturally a constant, and we simply have \begin{equation}
Z[V](\tau) = -24,
\end{equation}
where the number $24$ simply comes from the space of dimension-$1/2$ operators 
in the NS sector of the theory $F_{24}$ before the orbifolding,
and the sign is a convention which can be flipped by stacking the Arf theory.

From this theory $V$, a number of SCFTs with gravitational anomaly $-24n$
can be constructed by considering $V^n/S_n$, 
where $S_n$ acts by permuting $n$ copies of $V$.
Again, the supersymmetry makes $Z[V^n/S_n](\tau)$ a constant.
We can then consider the combination \begin{equation}
W_{n_1,\ldots,n_k} := \prod_{i=1}^k V^{n_i}/S_{n_i},
\end{equation} which would have a gravitational anomaly $-24^2$ if $\sum_i n_i = 24$.
It turns out that $Z_{W_{n_1,\ldots,n_k}}(\tau)$ does not in itself equal to $1$,
but we can take an integer linear combination of $W_{n_1,\ldots,n_k}$ 
of various choices of $n_i$ to obtain a theory $Y_-$ of gravitational anomaly $-24^2$ with $Z_{Y_-}(\tau)=1$.
This construction was first suggested by Gaiotto and Johnson-Freyd (unpublished),
and was carried out in detail in \cite{Albert:2022gcs}.

This is somewhat bothersome, since the theory $Y_-$ constructed in this way
is a direct sum of many different theories, and has multiple vacua in the NS sector.
Due to the importance of the periodicity element in the whole story, 
a more natural construction seems desirable.
Here comes another observation by Gaiotto and Johnson-Freyd (unpublished).
Recall that the theory $V$ has the symmetry $\Co_1$.
The anomaly of $\Co_1$ of this theory $V$ is known to be of order $24$ \cite{Johnson-Freyd:2017ble},
therefore the diagonal $\Co_1$ symmetry of $V^{24}$ is non-anomalous.
This suggests us to study the combined orbifold $V^{24}/[S_{24}\times \Co_1]$.
It turns out that there is a mixed anomaly between $S_{24}$ and $\Co_1$,
making this orbifold inconsistent.
This minor nuisance can be fixed by considering the orbifold $U:=V^{24}/[S_{24}\times \Co_0]$ instead.

This resulting theory has gravitational anomaly $-24^2$,
and the supersymmetry  makes its torus partition function with periodic spin structure a constant.
This `miraculous coincidence' of 
\begin{enumerate}
\item the appearance of the sporadic finite simple group $\Co_1$ after orbifolding $F_{24}$ by $\bZ_2$,
\item the fact that the anomaly of $\Co_1$ is of order $24$, and
\item the fact that the gravitational anomaly of the $24$ copies of $V$ so that the diagonal $\Co_1$ symmetry is non-anomalous is precisely equal to $-24^2$
which we expect for the periodicity element of TMF,
\end{enumerate}
made it very tempting to think that our theory $U$ is the desired SCFT $X_-$ of gravitational anomaly $-24^2$ with $Z(\tau)=1$.

Unfortunately, this expectation did not play out.
After developing the techniques explained in the rest of this paper
in the span of a couple of years, we found that \begin{equation}
Z[U](\tau) = 81270.
\end{equation}
This is by far the smallest absolute value
of the torus partition function with periodic spin structure
among the theories of gravitational anomaly $-24^2$
with single NS vacuum that people have found so far;
for example, \begin{equation}
Z[V^{24}/S_{24}](\tau) = 25499225,
\end{equation} as was determined in \cite{Albert:2022gcs}.

\subsection{Organization of the paper}
\label{intro:organization}

The rest of the paper is organized as follows.
In Sec.~\ref{sec:Conway-Duncan},
we review the necessary background on the Conway groups $\Co_0$ and $\Co_1$,
and the properties of the theory $V=F_{24}/\bZ_2$ first introduced by Duncan,
which has supersymmetry with a commuting action of $\Co_0$.
In particular, we explain how the $\Co_0$ symmetry acts on the Ramond sector of $V$,
and also report the structure of the commuting pairs $(g,h)$ of $\Co_0$ and $\Co_1$
we computed.
In Sec.~\ref{sec:PF}, 
we explain our computation of the torus partition function of the orbifold $V^{24}/[S_{24}\times \Co_0]$.
We start by explaining our choice of the orbifolding group.
We then proceed to discuss the methods to determine the phases $\alpha(g,h)$
for each commuting pair $(g,h)$ of the orbifolding group,
necessary to define the orbifold partition function.
We finally report the result of the computation.

We include one appendix, App.~\ref{app:ComputDetails}, 
where we provide more details of our computations.\footnote{
Some further details of the computation can be found in the PhD thesis of the second author \cite[Sec.~14.2]{Okada2025}.
Note that some notations there are not necessarily the same as those used here (e.g.\ the sign of \eqref{eq:def-of-Duncan-tw-pf-RR}),
and the description in the subsequent Sec.~14.3 contains some conjectural statements which are settled negatively in this paper.
}
We discuss how to obtain several group-theoretic data of the Conway groups $\Co_0,\Co_1$ in practice,
how to lift the embedding $\Co_0\subset SO(24)$ to $\Co_0\subset Spin(24)$,
how to implement the methods to compute the partition functions of $V^{24}$ and its orbifolds,
and how to expose the mixed anomaly between $\Co_1$ and $A_{24}$.
All the programs and data necessary to produce our results can be freely accessed at the third author's GitHub repository at 
\url{https://github.com/yujitach/conway-orbifold/}.

\section{Conway groups and Duncan's theory}
\label{sec:Conway-Duncan}
Let us start with a review of the Conway groups $\Co_0$ and $\Co_1$
and Duncan's theory $V$, which is the $\bZ_2$ orbifold
of the theory $F_{24}$ of 24 Majorana-Weyl fermions.

\subsection{Conway groups}
First we summarize basic facts about the Conway groups $\Co_0$ and
$\Co_1$ we will need in this paper.
We start from the standard information contained in \cite{ATLAS},
together with the group (co)homology data given e.g.~in \cite{Johnson-Freyd:2017bpn}.

There is a unique rank-24 even self-dual lattice $\Lambda\subset \bR^{24}$
which has no roots, i.e.~vectors of squared length $2$,
known as the Leech lattice.
The subgroup of $SO(24)$ fixing $\Lambda$ is
a finite group known as $\Co_0$,
whose center agrees with the center $\bZ_2$ of $SO(24)$: 
\begin{equation}
\bZ_2 \subset \Co_0 \subset SO(24).
\end{equation}
We denote the generator of the center by $z$.

The quotient of $\Co_0$ by this $\bZ_2$ is 
known as $\Co_1$:
\begin{equation}
1\to \bZ_2 \to \Co_0\to \Co_1 \to 1,\label{eq:Co0Co1ext}
\end{equation}
and this extension does not split.
The group $\Co_1$ is one of the 26 sporadic finite simple groups.
$\Co_0$ and $\Co_1$ are members of the Conway groups.
Their orders are \begin{align}
|\Co_0| &= 2^{22} 3^9 5^4 7^2  11 \cdot 13 \cdot 23 
\quad =\quad 8315553613086720000,\\
|\Co_1| &= 2^{21} 3^9 5^4 7^2  11 \cdot 13 \cdot 23 
\quad =\quad 4157776806543360000.
\end{align}

The group homologies are given by 
\begin{align}
H_1(B\Co_0;\bZ)&=1, &
H_2(B\Co_0;\bZ)&=0, &
H_3(B\Co_0;\bZ)&=\bZ_{24}, \label{eq:Co0homology}\\
H_1(B\Co_1;\bZ)&=1, &
H_2(B\Co_1;\bZ)&=\bZ_2, &
H_3(B\Co_1;\bZ)&=\bZ_{12}. \label{eq:Co1homology}
\end{align}
From this we see that \begin{equation}
H^2(B\Co_1;\bZ_2)=\bZ_2,
\end{equation}
meaning that there is a unique nontrivial extension
of $\Co_1$ by $\bZ_2$.
This extension is $\Co_0$, i.e.~the extension \eqref{eq:Co0Co1ext} 
is the unique nontrivial one.

$\Co_0$ has 167 conjugacy classes,
and $\Co_1$ has 101 conjugacy classes. 
This means that there are 66 pairs $[g]\neq[zg]$ 
of conjugacy classes of $\Co_0$,
where we remind the reader that $z$ is the generator of the center $\bZ_2\subset \Co_0$.
The rest, 35 conjugacy classes of $\Co_0$, are those such that 
$[g]=[zg]$.

Dually, $\Co_0$ and $\Co_1$ have 167 and 101 irreducible representations, respectively.
Any irreducible representation of $\Co_1$ is also an irreducible representation of $\Co_0$.
The rest, 66 irreducible representations of $\Co_0$,
are such that $z$ acts by $-1$.
In other words, they are genuine projective representations of $\Co_1$.

\subsection{Duncan's theory}

Let us consider the theory $F_{24}$ of 24 Majorana-Weyl fermions.
This has $SO(24)_1$ symmetry,
which acts projectively on the Ramond sector of the theory,
by acting on the spinor representations of $Spin(24)$.
That is, what acts genuinely on the Ramond sector is $Spin(24)_1$.
The lowest weight states of the Ramond sector have dimension $24/16=3/2$,
and they split into two chiral representations $S_+$ and $S_-$ of $Spin(24)$,
each of dimension $2^{12-1}=2048$.
We denote by $-1\in Spin(24)$
the element of the center which acts by $-1$ on both $S_\pm$.
The other two nontrivial elements of the center 
are the lift $\pm\hat z$ of the nontrivial center element $z$ of $SO(24)$.

We would like to consider the action of $\Co_0\subset SO(24)$ on the theory $F_{24}$.
Let $\hat \Co_0$ be the preimage of $\Co_0$ in $Spin(24)$,
which sits in the central extension sequence \begin{equation}
1\to \bZ_2 \to \hat \Co_0 \to \Co_0 \to 1.
\end{equation}
From \eqref{eq:Co0homology}, we see that $H^2(B\Co_0;\bZ_2)=0$,
meaning that this extension is trivial.
Therefore, there is a lift of $\Co_0 \subset SO(24)$
to $\Co_0 \subset Spin(24)$.
This lift is unique, since $H^1(B\Co_0;\bZ_2)=0$.

We already noted that there are two lifts $\pm\hat z$ to $Spin(24)$
of the central element $z\subset SO(24)$.
We denote simply by $z$ the one of $\pm \hat z$ which is in the lift of $\Co_0$.
This element $z$ acts by $+1$ on one of $S_\pm$ and by $-1$ on the other.
We fix the convention on the spinor representations 
so that it acts by $+1$ on $S_+$ and by $-1$ on $S_-$.
Then $S_+$ descends to a representation of $\Co_1=\Co_0/\langle z \rangle$,
while $S_-$ is a genuine projective representation of $\Co_1$.
It is known, see \cite[Proof of Prop.~4.4]{DuncanMackCrane} that $S_+$ decomposes 
as the representation of $\Co_1$ as\begin{equation}
S_+ = \mathbf{1} \oplus \mathbf{276} \oplus \mathbf{1776},
\end{equation}
where $\mathbf{1}$, $\mathbf{276}$, and $\mathbf{1776}$ are irreducible representations of $\Co_1$ of the indicated dimensions.
In particular, there is a unique $\Co_1$-invariant state in $S_+$.

Now, the NS sector and the R sector of $F_{24}$ have the following 
decompositions as the irreducible representation of $Spin(24)_1$:
\begin{align}
\cH_\text{NS}(F_{24}) &= [0] \oplus [V],&
\cH_\text{R}(F_{24}) &= [S_+] \oplus [S_-],
\end{align}
where $[0]$, $[V]$, $[S_\pm]$ are 
the vacuum, vector, and two spinor representations of $Spin(24)_1$,
respectively.
The $\bZ_2$ symmetry of $F_{n}$,
acting as $-1$ on all fermions, is anomaly free when $n$ is a multiple of $8$,
including our case $n=24$.
Let us then consider \begin{equation}
V:=F_{24}/\langle z\rangle,
\end{equation} where $z\in \Co_0$ is the generator of the center.
This acts by $+1$ on $[0]$ and $[S_+]$, and by $-1$ on $[V]$ and $[S_-]$.
Therefore, the NS and R sectors of $V$ are given by \begin{align}
\cH_\text{NS}(V) &= [0] \oplus [S_+],&
\cH_\text{R}(V) &= [V] \oplus [S_-].
\end{align}
This theory has $\Co_1$ symmetry,
which acts projectively on the R-sector.
The unique $\Co_1$-invariant state in $[S_+]$ 
is an operator of dimension $3/2$ in the NS sector of $V$
and generates a supersymmetry of the theory $V$,
as was first understood by Duncan and Mack-Crane \cite{Duncan,DuncanMackCrane}.

Note that the orbifold $V'=F_{24}/\langle -z\rangle$,
which is also a $\bZ_2$ orbifold of $F_{24}$ by the diagonal sign flip of all fermions,
is subtly different, and has the following decomposition of the NS and R sectors: \begin{align}
\cH_\text{NS}(V') &= [0] \oplus [S_-],&
\cH_\text{R}(V') &= [V] \oplus [S_+].
\end{align}
This theory has $\Co_0$ symmetry but not $\Co_1$ symmetry,
since the NS sector is not a genuine representation of $\Co_1$.
In this paper we stick to the theory $V$ with $\Co_1$ symmetry.

The anomaly of a fermionic theory in two dimensions with $G$ symmetry
is classified by the Pontryagin dual of the spin bordism group $\Hom(\Omega^\text{spin}_3(BG),U(1))$.
This can be computed from the group homology data given above,
as was done in \cite{Johnson-Freyd:2017ble}.
The result is that \begin{equation}
\Hom(\Omega^\text{spin}_3(B\Co_0),U(1)) 
= \Hom(\Omega^\text{spin}_3(B\Co_1),U(1)) 
= \bZ_{24},
\end{equation}
and the anomaly of Duncan's theory $V$ is a generator of this $\bZ_{24}$ \cite{Johnson-Freyd:2017ble}.
Therefore, the diagonal $\Co_1$ symmetry of $V^{24}$ is non-anomalous,
and we can consider the orbifold $V^{24}/\Co_1$.

We can also entertain the idea of considering the orbifolds
\begin{equation}
V^{24}/[S_{24}\times \Co_1]
\qquad\text{or}\qquad 
V^{24}/[A_{24}\times \Co_1],
\end{equation}
where the symmetric group $S_{24}$ or the alternating group $A_{24}$ acts by permuting 24 copies of $V$.
But these are possible only when there is no mixed anomaly between $S_{24}$ or $A_{24}$ and $\Co_1$.
We will see below that there is indeed a nontrivial mixed anomaly between $S_{24}$ or $A_{24}$ and $\Co_1$,
and we can only take
\begin{equation}
V^{24}/[S_{24}\times \Co_0]
\qquad\text{or}\qquad 
V^{24}/[A_{24}\times \Co_0].
\end{equation}
Before discussing why, 
we first need to understand the partition functions of $V$ with various holonomies of $\Co_1$.

\subsection{Torus partition functions of Duncan's theory}
\subsubsection{Free fermions and theta functions}
\label{sec:theta-conventions}
As Duncan's theory $V$ is a $\bZ_2$ orbifold of the theory $F_{24}$ of 24 Majorana-Weyl fermions,
we first need to understand the torus partition functions of $F_{24}$ with 
holonomies of $Spin(24)$.
It is well-known that they are given by products of theta functions
with characteristics,
but we need to carefully fix the conventions,
as the cochain-level expressions of anomalies crucially depend on the phases of the theta functions.

We start with our conventions for the theta functions.
We use $q=e^{2\pi i \tau}$ and $y=e^{2\pi i z}$,
and Jacobi's theta functions are defined by 
\begin{align}
\theta_3(\tau,z) & = \prod_{m=1}^\infty(1-q^m)(1+yq^{m-1/2})(1+y^{-1}q^{m-1/2}), \\
\theta_4(\tau,z) & = \prod_{m=1}^\infty(1-q^m)(1-yq^{m-1/2})(1-y^{-1}q^{m-1/2}), \\
\theta_2(\tau,z) & = q^{1/8}y^{1/2}\prod_{m=1}^\infty(1-q^m)(1+yq^{m})(1+y^{-1}q^{m-1}), \\
\theta_1(\tau,z) & = -\iu q^{1/8}y^{1/2}\prod_{m=1}^\infty(1-q^m)(1-yq^{m})(1-y^{-1}q^{m-1}). 
\end{align}
Using them, we define the partition functions of $F_2$,
i.e. the theory of one complex chiral fermion,
with holonomies $\lambda, \mu\in \bR$ 
in the convention where $\lambda,\mu=1$ corresponds to a $2\pi$ rotation of the fermion.
They are given by
\begin{align}
Z[F_2]{}_{\NS \lambda}^{\NS \mu}(\tau)
&:=  q^{\lambda^2/2}\frac{\theta_3(\tau,\lambda\tau+\mu)}{\eta(\tau)},&
Z[F_2]{}_{\NS \lambda}^{\rR \mu}(\tau)
&:=  q^{\lambda^2/2}\frac{\theta_4(\tau,\lambda\tau+\mu)}{\eta(\tau)}, \\
Z[F_2]{}_{\rR \lambda}^{\NS \mu}(\tau)
&:=  q^{\lambda^2/2}\frac{\theta_2(\tau,\lambda\tau+\mu)}{\eta(\tau)}, &
Z[F_2]{}_{\rR \lambda}^{\rR \mu}(\tau)
&:=  q^{\lambda^2/2}\frac{\theta_1(\tau,\lambda\tau+\mu)}{\eta(\tau)}, 
\end{align}
where 
the superscripts are for the temporal spin structure and holonomy,
and the subscripts are for the spatial spin structure and holonomy.
We then define
\begin{align}
Z[F_{2n}]{}_{\rX \vec{\lambda}}^{\rY \vec{\mu}}(\tau) := \prod_{i=1}^{n}Z[F_2]{}_{\rX(\lambda)_i}^{\rY(\mu)_i}(\tau) \ .
\label{eq:mult-cpx-ferm-XgYh-def}
\end{align} 
for the partition function of $2n$ Majorana-Weyl fermions with holonomies $\vec{\lambda},\vec{\mu}\in \bR^{n}$,
where $\rX,\rY\in\{\NS,\rR\}$. 

They have the following modular transformation properties.
First, under the $T$ transformation $\tau\mapsto \tau+1$, we have
\begin{align}
Z[F_2]{}_{\NS \lambda}^{\NS \mu}(\tau+1)
& = e^{-\pi\iu/12}e^{\pi\iu\lambda^2} Z[F_2]{}_{\NS \lambda}^{\rR (\lambda+\mu)}(\tau),\\
Z[F_2]{}_{\NS \lambda}^{\rR \mu}(\tau+1)
& = e^{-\pi\iu/12}e^{\pi\iu\lambda^2} Z[F_2]{}_{\NS \lambda}^{\NS (\lambda+\mu)}(\tau),\\
Z[F_2]{}_{\rR \lambda}^{\NS \mu}(\tau+1)
& = e^{\pi\iu/6}e^{\pi\iu\lambda^2} Z[F_2]{}_{\rR \lambda}^{\NS (\lambda+\mu)}(\tau),\\
Z[F_2]{}_{\rR \lambda}^{\rR \mu}(\tau+1)
& = e^{\pi\iu/6}e^{\pi\iu\lambda^2}  Z[F_2]{}_{\rR \lambda}^{\rR (\lambda+\mu)}(\tau).
\end{align}
Under the modular S transformation, 
we have
\begin{align}
Z[F_2]{}_{\NS \lambda}^{\NS \mu}(-\tfrac{1}{\tau})
& = e^{-2\pi\iu\lambda\mu} Z[F_2]{}_{\NS \mu}^{\NS(-\lambda)}(\tau),\\
Z[F_2]{}_{\NS \lambda}^{\rR \mu}(-\tfrac{1}{\tau})
& = e^{-2\pi\iu\lambda\mu} Z[F_2]{}_{\rR \mu}^{\NS(-\lambda)}(\tau),\\
Z[F_2]{}_{\rR \lambda}^{\NS \mu}(-\tfrac{1}{\tau})
& = e^{-2\pi\iu\lambda\mu} Z[F_2]{}_{\NS \mu}^{\rR (-\lambda)}(\tau),\\
Z[F_2]{}_{\rR \lambda}^{\rR \mu}(-\tfrac{1}{\tau})
& = -\iu e^{-2\pi\iu\lambda\mu} Z[F_2]{}_{\rR \mu}^{\rR (-\lambda)}(\tau).
\end{align}
Under the shifts $\lambda\mapsto\lambda+\frac{1}{2}$ or $\mu\mapsto\mu+\frac{1}{2}$,
we have
\begin{alignat}{2}
Z[F_2]{}_{\NS(\lambda+\frac{1}{2})}^{\NS\mu}(\tau) &= e^{-\pi\iu\mu} Z[F_2]{}_{\rR\lambda}^{\NS\mu}(\tau),
&\qquad Z[F_2]{}_{\NS\lambda}^{\NS(\mu+\frac{1}{2})}(\tau) &= Z[F_2]{}_{\NS\lambda}^{\rR\mu}(\tau), \\
Z[F_2]{}_{\NS(\lambda+\frac{1}{2})}^{\rR\mu}(\tau) &= \iu e^{-\pi\iu\mu} Z[F_2]{}_{\rR\lambda}^{\rR\mu}(\tau),
&\qquad Z[F_2]{}_{\NS\lambda}^{\rR(\mu+\frac{1}{2})}(\tau) &= Z[F_2]{}_{\NS\lambda}^{\NS\mu}(\tau), \\
Z[F_2]{}_{\rR(\lambda+\frac{1}{2})}^{\NS\mu}(\tau) &= e^{-\pi\iu\mu} Z[F_2]{}_{\NS\lambda}^{\NS\mu}(\tau),
&\qquad Z[F_2]{}_{\rR\lambda}^{\NS(\mu+\frac{1}{2})}(\tau) &= - Z[F_2]{}_{\rR\lambda}^{\rR\mu}(\tau), \\
Z[F_2]{}_{\rR(\lambda+\frac{1}{2})}^{\rR\mu}(\tau) &= \iu e^{-\pi\iu\mu} Z[F_2]{}_{\NS\lambda}^{\rR\mu}(\tau),
&\qquad Z[F_2]{}_{\rR\lambda}^{\rR(\mu+\frac{1}{2})}(\tau) &= Z[F_2]{}_{\rR\lambda}^{\NS\mu}(\tau).
\end{alignat}

\subsubsection{Duncan's theory as a $\bZ_2$ orbifold}

We can now write down the torus partition functions of Duncan's theory $V$
as follows: 
\begin{align}
Z[V]{}_{\NS\vec\lambda}^{\NS\vec\mu}(\tau)
& := \frac{1}{2}\Bigl( Z[F_{24}]{}_{\NS\vec\lambda}^{\NS\vec\mu}(\tau) + Z[F_{24}]{}_{\NS\vec\lambda}^{\rR\vec\mu}(\tau) + Z[F_{24}]{}_{\rR\vec\lambda}^{\NS\vec\mu}(\tau) + Z[F_{24}]{}_{\rR\vec\lambda}^{\rR\vec\mu}(\tau) \Bigr),
\label{eq:def-of-Duncan-tw-pf-NSNS} \\
Z[V]{}_{\NS\vec\lambda}^{\rR\vec\mu}(\tau)
& := \frac{1}{2}\Bigl( Z[F_{24}]{}_{\NS\vec\lambda}^{\NS\vec\mu}(\tau) + Z[F_{24}]{}_{\NS\vec\lambda}^{\rR\vec\mu}(\tau) - Z[F_{24}]{}_{\rR\vec\lambda}^{\NS\vec\mu}(\tau) - Z[F_{24}]{}_{\rR\vec\lambda}^{\rR\vec\mu}(\tau) \Bigr),
\label{eq:def-of-Duncan-tw-pf-NSR} \\
Z[V]{}_{\rR\vec\lambda}^{\NS\vec\mu}(\tau)
& := \frac{1}{2}\Bigl( Z[F_{24}]{}_{\NS\vec\lambda}^{\NS\vec\mu}(\tau) - Z[F_{24}]{}_{\NS\vec\lambda}^{\rR\vec\mu}(\tau) + Z[F_{24}]{}_{\rR\vec\lambda}^{\NS\vec\mu}(\tau) - Z[F_{24}]{}_{\rR\vec\lambda}^{\rR\vec\mu}(\tau) \Bigr),
\label{eq:def-of-Duncan-tw-pf-RNS} \\
-Z[V]{}_{\rR\vec\lambda}^{\rR\vec\mu}(\tau)
& := \frac{1}{2}\Bigl( Z[F_{24}]{}_{\NS\vec\lambda}^{\NS\vec\mu}(\tau) - Z[F_{24}]{}_{\NS\vec\lambda}^{\rR\vec\mu}(\tau) - Z[F_{24}]{}_{\rR\vec\lambda}^{\NS\vec\mu}(\tau) + Z[F_{24}]{}_{\rR\vec\lambda}^{\rR\vec\mu}(\tau) \Bigr).
\label{eq:def-of-Duncan-tw-pf-RR}
\end{align}
The relative signs in this set of four formulas are fixed, 
except for an overall sign ambiguity of the last one,
by the requirement that their modular transformation properties
are the same as those of $Z[F_{24}]{}_{\rX\vec\lambda}^{\rY\vec\mu}(\tau)$
including the anomalous phases associated to them.\footnote{%
Although the procedure of summing over $\bZ_2$ backgrounds might seem like 
we are summing over spin structures, we still end up having a fermionic theory
depending on the underlying spin structure of the worldsheet.
This happens when $c$ is an odd multiple of four.
For more details, see \cite{BoyleSmith:2024qgx}.
The method to choose the relative signs of four terms in the equations
\eqref{eq:def-of-Duncan-tw-pf-NSNS}
to 
\eqref{eq:def-of-Duncan-tw-pf-RR} is also explained in detail
in the same reference.
}
The sign in the last line is conventional. 
Flipping this sign to be $+1$ amounts to stacking the $\Arf$ theory, resulting in $\tilde V:=V\times \Arf$.

Now that we have carefully set up the conventions for the torus partition functions of Duncan's theory $V$
as functions of the holonomies $\vec\lambda,\vec\mu$ in $Spin(24)$,
we can now consider the torus partition functions of $V$ with holonomies 
restricted to lie in $\Co_1$.

Given a commuting pair $g,h\in \Co_1$,
lift them to $\hat g,\hat h\in \Co_0\subset Spin(24)$.\footnote{
    In practice, we first lift $g,h\in \Co_1$ to 
    $\tilde g,\tilde h\in \Co_0 \subset SO(24)$,
    and then lift $\tilde g,\tilde h\in SO(24)$
    to $\hat g,\hat h\in Spin(24)$.
    The first step is easily done, since elements of $\Co_1$
    are kept as elements of $SO(24)$ up to an overall sign
    in our computer implementation.
    The second step is trickier. $\tilde g$ and $\tilde h$
    have two lifts $\pm\hat g$ and $\pm\hat h$ in $Spin(24)$,
    and we need to pick the ones in $\Co_0$.
    This is done by picking the ones preserving
    the supercharge, which is the unique $\Co_1$-invariant 
    state in $S_+$.
    The lifting procedure is explained in more detail in Appendix~\ref{app:spinor}.
}
There are two cases: either $\hat g$ and $\hat h$ commute in $\Co_0$, 
or they satisfy $\hat g\hat h = z \hat h \hat g$ in $\Co_0$.
In the latter case, the holonomies $\hat g$ and $\hat h$
are actually inconsistent in the theory $F_{24}$ before the orbifolding,
and therefore the torus partition function of $V$ 
with holonomies $g$ and $h$ is simply zero.
In the former case, we can put $\hat g$ and $\hat h$ in the maximal torus of $Spin(24)$,
and represent them by $\vec\lambda,\vec\mu\in \bR^{12}$.

This construction of $(\vec\lambda,\vec\mu)$
from a commuting pair $(g,h)$ is not unique,
and a different choice of $(\vec\lambda,\vec\mu)$
produces a torus partition function differing by a phase,
which is a manifestation of the anomaly.
There is no canonical way to pick a representative $(\vec\lambda,\vec\mu)$ for each commuting pair $(g,h)$,
but we need to pick one. 
Once it is done, this needs to be kept fixed for all the subsequent computations.\footnote{%
The authors originally wanted to pick $\vec \lambda_g$ for each $g\in \Co_1$
and then to declare $(\vec \lambda,\vec \mu)$ for a commuting pair $(g,h)$ to be $(\vec \lambda_g,\vec \lambda_h)$.
But this did not work, as there turned out to be cases 
where two commuting pairs $(g,h)$ and $(g,k)$ sharing the same $g$ 
needed to be lifted to $(\lambda,\mu)$ and $(\lambda',\nu')$ 
with $\lambda\neq \lambda'$.
}
We are focusing on the torus partition function of $V$ with periodic spin structure, so we are interested in \begin{equation}
Z[V]{}_{\rR\vec\lambda}^{\rR\vec\mu}(\tau)
\end{equation}
for the commuting pairs $(g,h)$ of $\Co_1$.
When everything is set up correctly, the resulting function of $\tau$ is a constant,
which is a manifestation of the supersymmetry of the theory $V$.
This also provides a very useful nontrivial check of the correctness of the implementation of the above procedure.

In the rest of the paper, we only consider the torus partition function of $V$ with periodic spin structure.
Therefore, for brevity, we drop $\rR$ from the superscripts and the subscripts.

\subsection{Commuting pairs}
We have explained how we can compute the torus partition function of Duncan's theory. 
To make it concrete, we need to enumerate the commuting pairs of $\Co_1$. Because of a subtler reason related to the mixed anomaly,
we will also need to enumerate the commuting pairs of $\Co_0$.
At this point we are now in an uncharted territory,
where no ready-made information is available in the literature.
We used the computer algebra system GAP \cite{GAP4} to obtain the necessary data.
The details will be explained in Appendix~\ref{app:GAP},
and here we just summarize the results.

Let $G=\Co_1$ or $\Co_0$. 
We need to enumerate the conjugacy classes of commuting pairs $(g,h)$ of $G$,
where we say that $(g,h)$ and $(g',h')$ are conjugate 
if there is $k\in G$ such that $(g',h')=(kgk^{-1},khk^{-1})$.
We abbreviate the conjugacy class of $(g,h)$ by $[g,h]$,
and denote the set of conjugacy classes of commuting pairs by $X(G)$.

To determine $X(G)$, we first enumerate the conjugacy classes of $G$,
and pick a representative $g_i$ for each conjugacy class $[g_i]$.
We let $G_i=C_G(g_i)$ be the centralizer of $g_i$ in $G$.
Then we enumerate the conjugacy classes of $G_i$,
and pick a representative $h_j$ for each conjugacy class $[h_j]$ of $G_i$.
Then $[g_i,h_j]$ form a complete list of conjugacy classes of commuting pairs of $G$.
For later purposes, we actually need not just the conjugacy classes of $G_i$,
but the character tables of $G_i$ as well.

To analyze the modular transformation properties,
we also need to record the action of the modular group $SL(2,\bZ)$ on the conjugacy classes of commuting pairs. Namely, we need to compute the maps
\begin{equation}
T: [g,h]\mapsto [g,gh],\qquad
S: [g,h]\mapsto [h,g^{-1}].
\end{equation}
It turns out that it is also useful to record the action \begin{equation}
    P: [g,h]\mapsto [g,h^{-1}]
\end{equation} which extends $SL(2,\bZ)$ to $GL(2,\bZ)$,
and for $G=\Co_0$, the action of the central element $z$,
\begin{equation}
Z: [g,h]\mapsto [g,zh].
\end{equation}
We will also compute the map \begin{equation}
p: X(\Co_0) \to X(\Co_1)
\end{equation} given by 
\begin{equation}
[g,h] \mapsto [p(g),p(h)],
\end{equation}
where $X(G)$ denotes the set of conjugacy classes of commuting pairs of $G$,
and $p:\Co_0\to \Co_1$ is the projection map.

The computation for $\Co_1$ was doable on a laptop,
but the computation for $\Co_0$ required the use of a computer cluster.
The main issue was the memory requirement,
which was about 10GB for $\Co_1$ and about 170GB for $\Co_0$.
Going over conjugacy classes of $\Co_0$ and $\Co_1$ also took time,
and having a cluster with many nodes was very useful. 
We ended up using about 400 CPU hours in total.
We will explain the details of the computation in Appendix~\ref{app:GAP}.

\begin{figure}
\centering\includegraphics[width=\textwidth]{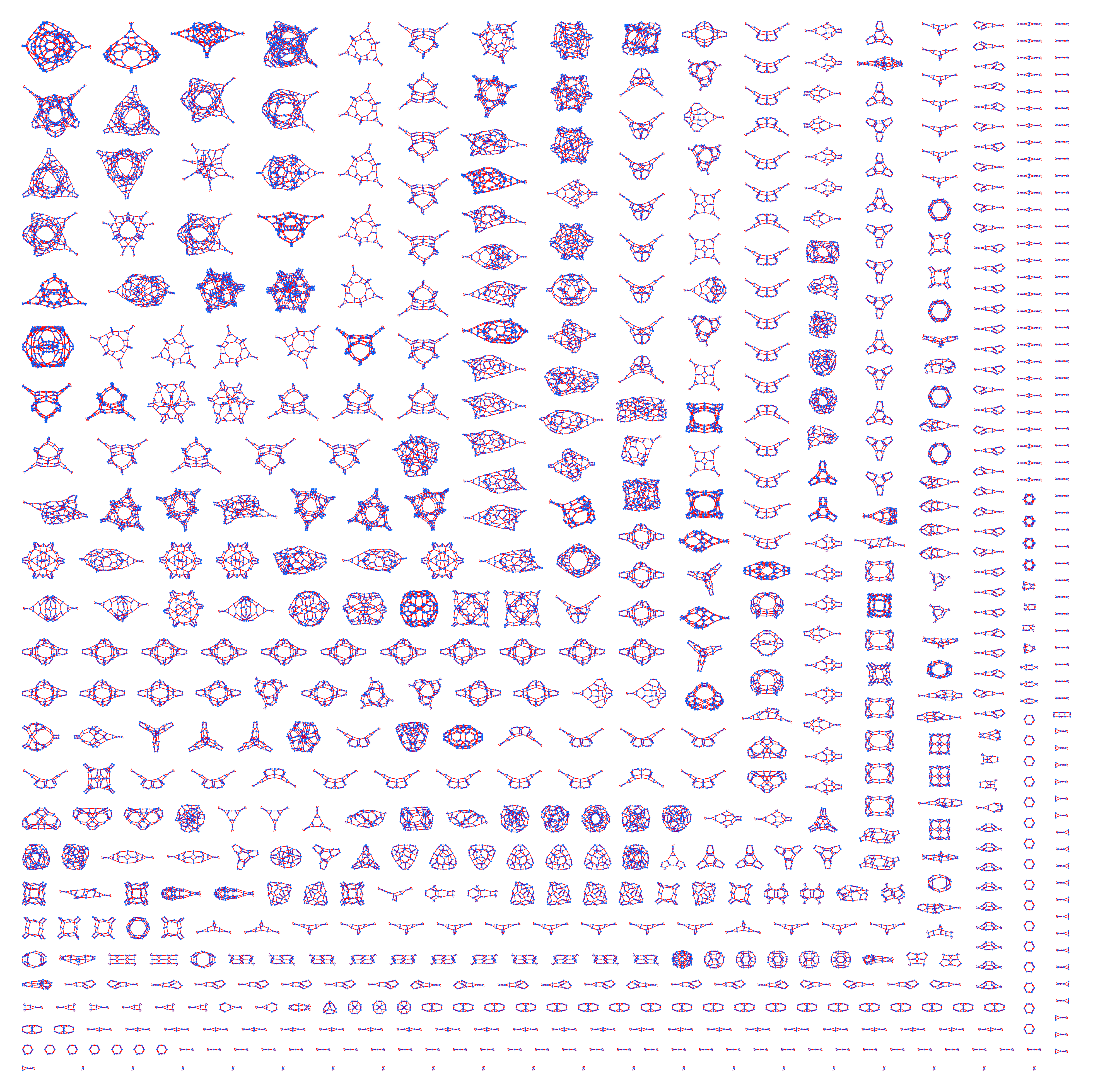}
\caption{The $SL(2,\bZ)$ orbits of 
conjugacy classes of commuting pairs of $\Co_0$.
Red and purple arrows indicate the action of $S$ and $T$, respectively.
\label{fig:co0-orbits}
}
\end{figure}
\begin{figure}
\centering\includegraphics[width=\textwidth]{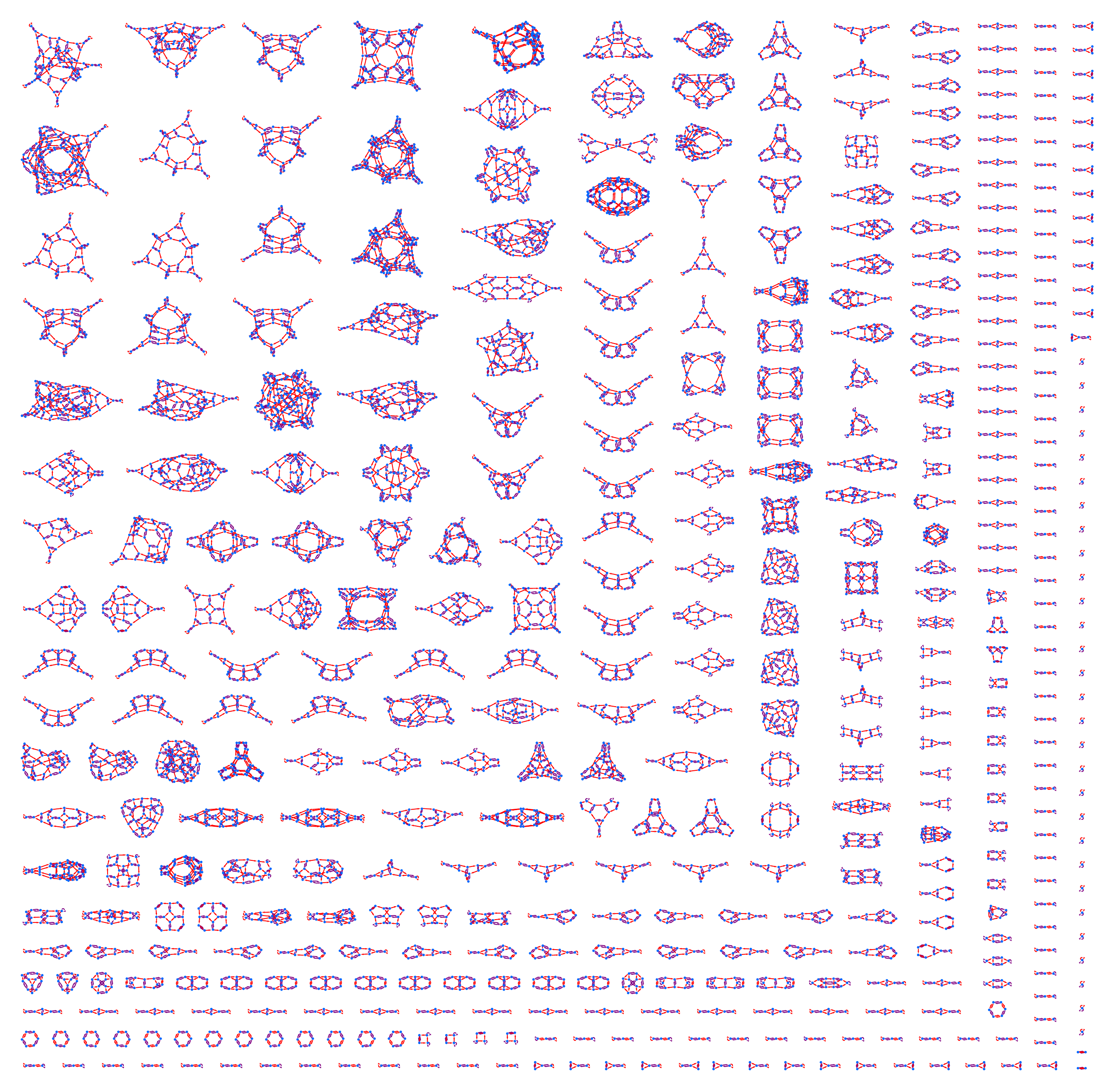}
\caption{The $SL(2,\bZ)$ orbits of 
conjugacy classes of commuting pairs of $\Co_1$.
Red and purple arrows indicate the action of $S$ and $T$, respectively.
\label{fig:co1-orbits}
}
\end{figure}

After the dust settled, we found that the numbers of 
conjugacy classes of commuting pairs are \begin{equation}
|X(\Co_1)| = 7578, \qquad |X(\Co_0)| = 22017,
\end{equation} while the number of
$SL(2,\bZ)$ orbits of $X(\Co_1)$ and $X(\Co_0)$ are \begin{equation}
\left|\frac{X(\Co_1)}{SL(2,\bZ)}\right| = 445, \qquad
\left|\frac{X(\Co_0)}{SL(2,\bZ)}\right| = 699.
\end{equation}
The $SL(2,\bZ)$ orbits of $X(\Co_0)$ and $X(\Co_1)$ are shown in Figures~\ref{fig:co0-orbits} and \ref{fig:co1-orbits}, respectively.

Using the data of the conjugacy classes of commuting pairs,
we can compute the torus partition functions of Duncan's theory $V$ with holonomies in $\Co_1$.
We found that the torus partition functions are indeed always constant.
They also all agreed with the results of Johnson-Freyd and Furet \cite{Johnson-Freyd:2023uvb},
where the torus partition functions of $V$ with only spatial holonomies were computed.

\section{Partition functions of Conway orbifolds}
\label{sec:PF}
\subsection{Orbifolds we consider}
As mentioned above,
the $\Co_1$ anomaly of $V^{24}$ is trivial,
and therefore we can consider the orbifold $V^{24}/\Co_1$.
We can also consider $V^{24}/\Co_0$,
but the center $\bZ_2$  of $\Co_0$ acts trivially on the NS sector 
of $V^{24}$, and therefore the orbifold $V^{24}/\Co_0$ is equivalent to 
two decoupled copies of $V^{24}/\Co_1$:
\begin{equation}
V^{24}/\Co_0 \cong V^{24}/\Co_1 \sqcup V^{24}/\Co_1.
\end{equation}

It is also tempting to consider the orbifolds $V^{24}/[S_{24}\times \Co_1]$ or $V^{24}/[A_{24}\times \Co_1]$,
where $S_{24}$ or $A_{24}$ acts by permuting 24 copies of $V$.
Unfortunately, there is a nontrivial mixed anomaly between $S_{24}$ or $A_{24}$ and $\Co_1$,
and therefore these orbifolds are not well-defined.
Instead, we can consider the orbifolds $V^{24}/[S_{24}\times \Co_0]$ or $V^{24}/[A_{24}\times \Co_0]$,
which are guaranteed to be well-defined,
as we will argue below.
We will also see below that 
$V^{24}/[S_{24}\times \Co_0]$ has a single NS-sector vacuum,
while $V^{24}/[A_{24}\times \Co_0]$ has two NS-sector vacua,
and splits into two copies of $V^{24}/[S_{24}\times \Co_0]$:
\begin{equation}
V^{24}/[A_{24}\times \Co_0] \cong V^{24}/[S_{24}\times \Co_0] \sqcup V^{24}/[S_{24}\times \Co_0].
\end{equation}

As we already mentioned, we can also study the theory $\tilde V=V\times \Arf$, obtained by stacking
with the Arf theory. This flips the overall sign of \eqref{eq:def-of-Duncan-tw-pf-RR}.
We can then consider $\tilde V^{24}/[S_{24}\times \Co_0]$.
The corresponding alternating orbifold splits into two copies as before.

Finally, we note that when we perform $A_{24}$ or $S_{24}$ orbifolding,
there is a freedom to include a $\bZ_2$ discrete torsion phase.
In this paper we will not consider this generalization.
The authors hope to revisit this question in a future publication.

\subsubsection{Mixed anomaly between $\Co_1$ and $S_{24}$}
\label{sec:mixed-S24}
Let us start with the mixed anomaly between $\Co_1$ and $S_{24}$,
which is straightforward to identify.
We consider the element $\pi\in \bZ_{24}\subset S_{24}$ cyclically rotating the 24 copies of $V$,
and consider the $\pi$-twisted NS sector.
This is the same as the theory $V$ put on a circle which is a 24-fold cover of the original spatial circle.
This is in the R-sector, and therefore the lowest-energy states form $\bR^{24}$, 
on which $\Co_1$ acts projectively
as $\Co_0$.
Therefore, $\Co_1$ and $S_{24}$ have mixed anomaly.

\subsubsection{Mixed anomaly between $\Co_1$ and $A_{24}$}
\label{sec:mixed-A24}

The argument above does not work for the mixed anomaly between $\Co_1$ and $A_{24}$, 
since $\pi$ is not in $A_{24}$. 
We need a more complicated argument.
The rest of this subsubsection is very technical, 
and the reader may skip it on a first reading.

Let us regard $V^{24} = W_1 W_2 W_3 W_4$, where each of $W_i\simeq V^6$ 
consists of six copies of $V$.
Let $c_i$ be the cyclic shift by 1 of six copies of $V$ inside $W_i$, 
and $a_i=c_i^3$ be the shift by 3. Note that $a_i^2=e$.
Let $b_{12}$ be the swap of two factors $W_1$ and $W_2$, and 
similarly for $b_{34}$.
Now, $a_1$, $a_2$ and $b_{12}$ generate a $D_8$ (the dihedral group with eight elements).
$a_3$, $a_4$ and $b_{34}$ similarly generate another copy of $D_8$.
Let $G_S=D_8\times D_8 \subset S_{24}$.
An index-2 subgroup $G_A:=G_S\cap A_{24}\subset G_S$ (with 32 elements) is a subgroup of $A_{24}$.

Let $c:=c_1c_2c_3c_4 \subset A_{24}$. 
Note that $G_A$  commutes with $c$.
Take the commuting pair $(g,h)$ of $\Co_1$ called $\#(4,21)$ 
in the notation of Appendix~\ref{app:GAP}.
The ATLAS names of $g$ and $h$ are 26A and 2B, respectively,
and their lifts $\hat g$, $\hat h$
in $\Co_0$ satisfy $\hat g\hat h = z \hat h \hat g$.

We consider the $gc$-twisted sector, and derive a contradiction assuming that 
$G_A$ and $h$ act non-projectively. 
Note that $h$ is of order two, so the trace of $h$ in any representation should be an integer.
So, it suffices to show that the trace of $h$ in the $G_A$-invariant part of $gc$-twisted sector is not an integer.
This is given by \begin{equation}
X=\frac{1}{32}\sum_{x\in G_A} Z[V^{24}]_{gc}^{xh}.
\end{equation}
Each term is given by products of various $Z[V]_{g^k}^{h^i g^j}$ for various $i,j,k$,
(up to a phase).
Importantly, this turns out to be zero unless $i$, $j$ and $k$ are all even.\footnote{%
The authors do not know a conceptual reason for this,
but it can be checked explicitly as follows,
by separately studying the even-$i$ case and the odd-$i$ case.
\begin{itemize}
\item When $i$ is even, $h^2=e$, so the pair $(h^i g^j, g^k)$ 
is $SL(2,\bZ)$-equivalent to $(e,g^{\gcd(j,k)})$.
Assuming not all of $i$,$j$ and $k$ are even, $\gcd(j,k)$ is odd,
meaning that $g^{\gcd(j,k)}$ is in the conjugacy class 26A of $\Co_1$.
Its lift to $\Co_0$ is in 52A.
So we need to look up $Z[V]_e^\text{52A}$ in our pre-computed table.
We find that it is zero.
\item Next we consider the case when $i$ is odd.
When  $k$ is odd, the partition function  vanishes,
since the lifts of $h^i g^j$ and $g^k$ to $\Co_0$ do not commute.
When $k$ is even, 
we can use GAP to check that the pair $(h^i g^j, g^k)$
is $SL(2,\bZ)$-equivalent to either $(e,h)$ or $(e,g)$.
$h$ is in 2B of $\Co_1$, which lifts to 4A of $\Co_0$,
and $g$ is in 26A of $\Co_1$, which lifts to 52A of $\Co_0$.
So we need to look up $Z[V]_e^\text{4A}$ and $Z[V]_e^\text{52A}$ in 
our pre-computed table.
We find that both of them are zero.
\end{itemize}
For more details, see Appendix~\ref{app:mixed}.
}

Now, to make the power of $h$ to be even, $x$ needs to contain both $b_{12}$ and $b_{34}$,
so we have \begin{equation}
X= \frac{1}{32} \sum_{i=0,1}\sum_{j=0,1} Z[V^{24}]_{gc}^{b_{12}b_{34} (a_1a_2)^i (a_3a_4)^j h} .
\end{equation}
Now, $b_{12}$ and $b_{12}a_1a_2$ are conjugate, because $a_1 b_{12}=b_{12}a_2$.
Similarly, $b_{34}$ and $b_{34}a_3a_4$ are conjugate.
As the conjugate elements have the same trace in any genuine representation,
all four terms have the same value, and we have \begin{equation}
X=\frac18 Z[V^{24}]_{gc}^{b_{12}b_{34}h} \propto \frac18 (Z[V]_{g^{6}} )^2
\end{equation} where the $\propto$  sign means an equality up to a complex phase.
As $Z[V]_{g^{6}}\propto 2$, \begin{equation}
X\propto \frac12,
\end{equation}
which is a contradiction.

\subsubsection{$\Co_0$ symmetries of $V^{24}/S_{24}$ and $V^{24}/A_{24}$}
\label{sec:our-orbifolds}

Instead, we consider the orbifolds $V^{24}/[S_{24}\times \Co_0]$ and $V^{24}/[A_{24}\times \Co_0]$.
To do this, we first need to show that
$V^{24}/S_{24}$ and $V^{24}/A_{24}$ have non-anomalous $\Co_0$ symmetry.
This can be argued as follows.
$V^{24}/S_{24}$ and $V^{24}/A_{24}$ 
both have dimension-1 currents generating $\mathfrak{spin}(24)_{24}$ 
affine algebra.
This integrates to an affine $Spin(24)_{24}$ symmetry of the theory.
We now take the subgroup $\Co_0\subset Spin(24)$.
Its anomaly is 24 times the anomaly of $\Co_0$ 
pulled back from $Spin(24)_{1}$ symmetry.
As the space of fermionic anomalies of $\Co_0$ is $\bZ_{24}$,
we conclude that the $\Co_0$ symmetry of $V^{24}/S_{24}$ and $V^{24}/A_{24}$ is non-anomalous.

Now, the central element $z\in \Co_0$ acts trivially on the NS sector of $V$.
This raises the question if $z$ acts trivially on the NS sector of $V^{24}/S_{24}$ and $V^{24}/A_{24}$,
creating two dimension-0 operators in the NS sector of these theories.
To study this issue, 
we perform the orbifolds $V^{24}/[S_{24}\times \Co_0]$ and $V^{24}/[A_{24}\times \Co_0]$ in the opposite order,
first by $\Co_0$ and then by $S_{24}$ or $A_{24}$.
Now, $V^{24}/\Co_0$ has two dimension-0 operators in the NS sector,
and the question is whether $S_{24}$ or $A_{24}$ acts trivially 
or non-trivially on them.

Note that the two dimension-0 operators in the NS sector of $V^{24}/\Co_0$ 
correspond to two projectors $e_+$ and $e_-$ satisfying
$e_+ + e_- = 1$ and $e_+ e_- = 0$.
Therefore, there should be a permutation action of $S_{24}$ and $A_{24}$
on the two projections.

Group theory tells us that the action of $S_{24}$ 
on the two projectors is either the trivial one,
or the non-trivial one where odd permutations swap $e_+$ and $e_-$.
This already means that $A_{24}$ acts trivially on the two projectors.

To decide which of the two possibilities is realized for $S_{24}$,
we consider the odd permutation $\pi\in S_{24}$ cyclically rotating the 24 copies of $V$,
and study how $z$ acts on the $\pi$-twisted NS sector of $V^{24}$.
The lowest-weight state of the $\pi$-twisted NS sector of $V^{24}$ is
simply $\bR^{24}$, on which $z$ acts by $-1$.
Using the modular invariance to regard this phase
as arising from the action of $\pi$ in the $z$-twisted NS sector,
we conclude that odd permutations of $S_{24}$ swap the two projectors.
This is actually the same analysis as in Sec.~\ref{sec:mixed-S24}
(now applied to the NS sector).

From this we conclude the following:
$V^{24}/[A_{24}\times \Co_0]$ has two vacua,
arising from the trivial action of $\bZ_2\subset \Co_0$.
Then, $\bZ_2\simeq S_{24}/A_{24}$
acts non-trivially by swapping the two vacua, and therefore
$V^{24}/[S_{24}\times \Co_0]$ has only one vacuum.
Moreover, this means that we have the decomposition \begin{equation}
V^{24}/[A_{24}\times \Co_0] \cong V^{24}/[S_{24}\times \Co_0] \sqcup V^{24}/[S_{24}\times \Co_0].
\end{equation}

\subsection{Consistency conditions on anomaly cancelling phases}
So far we introduced the orbifold theories we consider.
They are $\Co_1$ and $\Co_0$ orbifolds of 24 copies of $V$, 
which are \begin{equation}
V^{24}/\Co_1, \qquad V^{24}/\Co_0\simeq V^{24}/\Co_1 \sqcup V^{24}/\Co_1,
\end{equation} together with the symmetric and alternating orbifolds,
which are \begin{equation}
V^{24}/[S_{24}\times \Co_0]
\end{equation} and \begin{equation}
V^{24}/[A_{24}\times \Co_0]\simeq V^{24}/[S_{24}\times \Co_0] \sqcup V^{24}/[S_{24}\times \Co_0].
\end{equation}
We can also replace $V$ with $\tilde V=V\times \Arf$.
This does not change $V^{24}/\Co_1$ or $V^{24}/\Co_0$,
but it does affect the symmetric and alternating orbifolds, since the minus sign introduced by stacking the Arf theory
affects the twisted sectors.

Note that $V^{24}$, $V^{24}/S_{24}$ and $V^{24}/A_{24}$ 
all have $Spin(24)_{24}$ symmetry, and therefore $\Co_0$ symmetry.
As we argued, this $\Co_0$ symmetry is non-anomalous,
therefore we can consider the orbifolds $V^{24}/\Co_0$,
$(V^{24}/S_{24})/\Co_0$ and $(V^{24}/A_{24})/\Co_0$,
and these are exactly the theories we are interested in.
We do know that the third one splits 
into two copies of the second one,
but the computations we will perform will not directly see
this splitting, 
and confirming this doubling gives us a nontrivial check of the computations.

Now consider a general question.
Let $T$ be a theory with $Spin(24)_{24}$ symmetry.
We fix the conventions for the torus partition functions of $T$
with holonomies in $Spin(24)$.
In particular, we fix the anomalous phases appearing 
in the modular transformation properties of the torus partition functions,
which are independent of the theory $T$
and are determined by the $Spin(24)_{24}$ symmetry alone.
Let $Z[T]_{\vec\lambda}^{\vec\mu}(\tau)$ be
the torus partition function of $T$ with periodic spin structure
and with holonomies $\vec\lambda,\vec\mu\in\bR^{12}$.

We now take the subgroup $\Co_0\subset Spin(24)$.
Then $\Co_0$ is non-anomalous.
We also assume that $T$ has a supercharge invariant under $\Co_0$.
Then $Z[T]_{\vec\lambda}^{\vec\mu}(\tau)$ is a constant, 
determined by $(\vec\lambda,\vec\mu)$ associated to a commuting pair $(g,h)$ of $\Co_0$.

The anomaly of the $\Co_0$ symmetry is guaranteed to be trivial,
but the anomaly cocycle will be a nonzero coboundary.
This means that the torus partition function $\hat Z[T]_g^h$
which is manifestly modular covariant 
is different from $Z[T]_{\vec\lambda}^{\vec\mu}$
by a phase, that is, we have
\begin{equation}
\hat Z[T]_g^h = \alpha(g,h) Z[T]_{\vec\lambda}^{\vec\mu},
\label{alphadef}
\end{equation}
where $\alpha(g,h)$ is a $U(1)$-valued function of commuting pairs $(g,h)$ of $\Co_0$,
and the pair of $Spin(24)$ holonomies $(\vec\lambda,\vec\mu)$ is
chosen and uniquely fixed for each $(g,h)$ as explained above.
We know how to compute $Z[T]_{\vec\lambda}^{\vec\mu}$ once we 
do and did fix the conventions,
but we do not know how to compute $\alpha(g,h)$ from first principles.
We call these unknown phases $\alpha(g,h)$ the anomaly-cancelling phases.

We will determine $\alpha(g,h)$ indirectly,
using the following four requirements: 
\begin{enumerate}
\item $\hat Z[T]_g^h$ should be modular covariant without any additional phase,
\item there should be a relation $\hat Z[T]_g^h=\tr_{\cH_g[T]} (-1)^F h$ 
where $\cH_g[T]$ is a vector space graded by $(-1)^F$
on which $C_{\Co_0}(g)$ acts in a genuine representation,
\item the phases $\alpha(g,h)$ are universal and do not depend
on the theory $T$, 
\item and $\alpha(g,h)$ should take values in a finite set of values
allowed by the anomaly analysis of the Abelian subgroup $\langle g,h\rangle\subset \Co_0$,
where $\langle g,h\rangle$ is the subgroup
generated by $g$ and $h$.
\end{enumerate}
Let us discuss these four points in detail,
one in each subsubsection.

\subsubsection{Modular covariance}

In a non-anomalous theory $T$, the partition function $\hat Z[T]_g^h(\tau)$
on the torus with periodic spin structure and with holonomies $g$ and $h$
should be modular covariant without any additional phase,
namely we should have \begin{equation}
\hat Z[T]_g^h(-\tfrac{1}{\tau}) = \hat Z[T]_h^{g^{-1}}(\tau),\qquad
\hat Z[T]_g^h(\tau+1) = \hat Z[T]_g^{gh}(\tau)
\end{equation}
for the actions of $S$ and of $T$, respectively.
This gives the $SL(2,\bZ)$ action on the set of $\hat Z[T]_g^h$.

In our supersymmetric case,
no $\tau$ dependence is present, and the relations become simpler.
In practice, what we do is the following.
For each $SL(2,\bZ)$ orbit $\cO$ of commuting pairs of $\Co_0$,
we pick a representative $(g_\cO,h_\cO)$ and study $\hat Z[T]_g^h$.
This value can then be reused for any other commuting pair in the same orbit.
As there are 699 $SL(2,\bZ)$ orbits of commuting pairs of $\Co_0$,
this means that we need to determine $\alpha(g,h)$ for
only 699 commuting pairs.

We will also use the following related property.
Note that in any representation $\chi$ of a finite group $G$,
we have
\begin{equation}
\tr_\chi g^{-1} = \overline{\tr_\chi g}.
\end{equation}
Combined with the fact that the $L_0$ eigenvalues are real
in a unitary theory, we can derive the following relation
which holds in any unitary theory:
\begin{equation}
\overline{\hat Z[T]_g^h(\tau)} = \hat Z[T]_g^{h^{-1}}(-\bar \tau).
\label{eq:Prelation}
\end{equation}
We call this the $P$ action,
which is a consequence of CPT symmetry of unitary theories.
This extends the $SL(2,\bZ)$ action previously mentioned 
to a $GL(2,\bZ)$ action.

In our case, $\tau$ dependence is absent. 
When $P\cO=\cO$, the relation \eqref{eq:Prelation} means 
that $\hat Z[T]_g^h$ is real. 
This only leaves two possible choices for $\alpha(g,h)$.\footnote{%
Strictly speaking, this only applies when $\hat Z[T]_g^h$ is nonzero.
But as we argue below, $\alpha(g,h)$ is independent of the theory $T$,
and we only need a single theory $T$ for which $\hat Z[T]_g^h$ is nonzero
to conclude this reduction of possibilities.
It turns out by explicit computations that there are indeed theories $T$ 
for which $\hat Z[T]_g^h$ is nonzero for all 699 orbits.
So indeed there are only two possible choices for $\alpha(g,h)$ for each $P$-invariant orbit.
}
We write one such choice by $\alpha_0(g,h)$ and parameterize the choices
by writing \begin{equation}
\alpha(g_\cO,h_\cO) = A_\cO \alpha_0(g_\cO,h_\cO),\qquad A_\cO=\pm1.
\end{equation}

In contrast, when $P\cO\neq \cO$, 
this relates $\hat Z[T]_g^h$ and $\hat Z[T]_g^{h^{-1}}$,
and therefore reduces the number of independent choices of $\alpha(g,h)$.
In these cases, we choose the representatives $(g_\cO,h_\cO)$ and $(g_{P\cO},h_{P\cO})$ such that $g_\cO=g_{P\cO}$ and $h_\cO=h_{P\cO}^{-1}$.
Furthermore, we pick the angle vectors in $\bR^{12}$ representing the commuting pairs for $\cO$ and $P\cO$  
such that $(\vec\lambda_{P\cO},\vec\mu_{P\cO})=(\vec\lambda_{\cO},-\vec\mu_{\cO})$.
Then we use \begin{equation}
B_\cO:=\alpha(g_\cO,h_\cO),\qquad
B_{P\cO}:=\alpha(g_{P\cO},h_{P\cO}).
\end{equation}
to parameterize the choices of $\alpha(g,h)$ associated to the two orbits $\cO$ and $P\cO$. 
With these choices, $Z[T]_{\vec \lambda}^{-\vec \mu} = \overline{Z[T]_{\vec \lambda}^{\vec \mu}}$,
therefore $B_{P\cO}=\overline{B_\cO}$,
giving us a convenient parameterization for the anomaly-cancelling phases
associated to these two orbits.

It turns out that, out of 699 orbits,
639 orbits are $P$-invariant, and the remaining 60 orbits are paired by the $P$ action.
Therefore we only need to determine 639 signs $A_{\cO}$
and 30 complex phases $B_{\cO}$.

\subsubsection{Correct decomposition into irreducibles}

To further constrain the anomaly-cancelling phases $\alpha(g,h)$,
we require that $\hat Z[T]_g^h$ should be interpreted as the trace of $h$ in a vector space $\cH_g[T]$ graded by $(-1)^F$ on which $C_{\Co_0}(g)$,
the centralizer of $g$ in $\Co_0$, acts in a genuine representation. 
If this would be the case, the multiplicity  of an irreducible representation $\chi$ of $C_{\Co_0}(g)$ in $\cH_g[T]$,
again weighted by $(-1)^F$, should be given by  \begin{equation}\label{eq:n[T]}
    n[T](g,\chi):=\sum_{[h]}\frac{\#[h]}{|C_{\Co_0}(g)|}  \bar\chi(h) \hat Z[T]_g^h,
\end{equation}
where the sum is over the conjugacy classes $[h]$ of $C_{\Co_0}(g)$, 
$\#[h]$ is the number of elements in the conjugacy class $[h]$,
and we identified an irreducible representation and its character for brevity of notation.

Our requirement is then that $n[T](g,\chi)$ defined above should be integers for all $g,\chi$.
Using \eqref{alphadef} and the modular covariance,
this gives a set of constraints of the form 
\begin{equation}
\sum_{\cO=P\cO} z[T](g,\chi)_\cO A_{\cO} + 
\sum_{\cO\neq P\cO} w[T](g,\chi)_\cO B_{\cO} \in \bZ
\label{ABeq}
\end{equation}
for each conjugacy class $[g]$ of $\Co_0$ and each irreducible representation $\chi$ of $C_{\Co_0}(g)$,
where $z[T](g,\chi)_\cO$ and $w[T](g,\chi)_\cO$ are complex numbers determined by the partition function $Z[T]_{\vec\lambda}^{\vec\mu}$ of the theory $T$ for the angles $(\vec\lambda,\vec\mu)$ of the orbit representative $(g_\cO,h_\cO)$.
This gives a set of strong constraints on the anomaly-cancelling phases $A_\cO$ and $B_\cO$.

\subsubsection{Universality of anomaly-canceling phases}

The next constraint is that the anomaly-cancelling phases $\alpha(g,h)$ should be universal and independent of the theory $T$,
as long as we use the same set of conventions 
for the torus partition functions of 
$T$ as a theory with level-$24$ $Spin(24)$ symmetry.
This is because $\alpha(g,h)$ should be in principle computable 
from the anomaly cochain of $\Co_0$ symmetry of $T$,
which should then be determined by the level-$24$ $Spin(24)$ symmetry alone.

In practice, this means that we can impose the integrality condition \eqref{ABeq} for multiple theories $T_{1,\ldots}$,
to reduce the possibilities of the anomaly-cancelling phases $A_\cO$ and $B_\cO$.
Once we have enough theories $T_{1,\ldots,k}$ to reduce the possibilities to a single choice,
the integrality condition \eqref{ABeq} for any other theory $T'$ should be 
automatically satisfied.
This gives a strong consistency check of the whole procedure.

\subsubsection{Consistency when restricted to $\langle g,h\rangle$}
\label{sec:abelianDefault}
We noted above that, for orbits $\cO$ with $P\cO=\cO$,
the anomaly-cancelling phases $A_\cO$ are $\pm1$.
But so far we only know that, for orbits $\cO$ with $P\cO\neq\cO$,
 $B_\cO$ is a complex number of absolute value $1$.
We would like to make a further restriction on $B_\cO$
so that the number of possibilities is finite.

For this, we consider the subgroup $G:=\langle g,h\rangle\subset \Co_0$ generated by $g$ and $h$. 
As $g$ and $h$ commute, this is an Abelian subgroup.
The set of anomaly-cancelling phases of $\Co_0$
should restrict to a set of anomaly-cancelling phases of this subgroup.
As Abelian groups are much simpler,
it is possible to develop a complete theory of anomaly-cancelling phases for them.
We will leave this to a future work, and will state only 
the minimal amount of information necessary for our purposes.

Without loss of generalities, by choosing a suitable representative of the $SL(2,\bZ)$ orbit,
we can assume that $\langle g,h\rangle\cong \bZ_a\times \bZ_b$
where $g$ and $h$ generate $\bZ_a$ and $\bZ_b$, respectively,
and $a$ divides $b$.

For the moment, let us consider a general theory $T$ with 
$G=\langle g,h\rangle$ symmetry as above,
where $g$ and $h$ are embedded in some $(U(1)_1)^n$ symmetry.
We pick explicit angles $\vec\mug,\vec \muh \in \bR^n$ representing the elements $g$ and $h$.\footnote{%
Note that so far our convention was that 
$\vec\lambda$ and $\vec\mu$ associated to the commuting pair $(g,h)$ in $\Co_0$ take values in 
$\bR^{12}$, appropriate for a single copy of the Conway theory $V$, with $Spin(24)_{1}$  symmetry.
The analysis here applies to 24 copies of $V$, with $Spin(24)_{24}$ symmetry.
This means that $\vec\mug$ and $\vec\muh$ here take values in $\bR^{24\times 12}$.
We will come back to this issue in \eqref{foofinal}.
}
We further assume that the anomaly cocycle of $G$ is bosonic.

Under these assumptions, the information we use is that 
\begin{equation}
\hat Z_{mg+nh}^{m'g+n'h} := e^{\pi i mm' \vec\mug\cdot\vec\mug +2\pi imn'\vec\mug \cdot \vec\muh + \pi i nn'\vec\muh\cdot \vec\muh}Z_{m\vec\mug+n\vec\muh}^{m'\vec\mug+n'\vec\muh}
\label{foo}
\end{equation}
is \emph{a} consistent anomaly-free choice of twisted partition function as a $\langle g,h\rangle$-symmetric theory.
Restricting to $(m,n;m',n')=(1,0;0,1)$, this means that \begin{equation}
\alpha(g,h)=e^{2\pi i \vec\mug\cdot\vec\muh}
\end{equation}
is \emph{a} consistent choice of the anomaly-cancelling phases.\footnote{%
Note that this statement depends on our conventions for the torus partition functions set up in Sec.~\ref{sec:theta-conventions},
and cannot be applied to other conventions without modification.
Also, when $a$ is not chosen to divide $b$,
the formula needs a modification; see also the next footnote.
}

The justification of this statement is as follows.
In our conventions set up in Sec.~\ref{sec:theta-conventions},
the partition function of a theory with $U(1)_1$ symmetry
in the  $\lambda$-twisted sector with the temporal twist parameter $\mu$
is defined to contain only (half-)integral powers of $y$.
In the Hamiltonian language, this means that we chose \emph{not}
to include the fractional $\lambda$-dependent $U(1)$ charge,
which would have contributed an additional phase in the partition function.
In the above equation \eqref{foo}, 
we judiciously provide the missing phases, depending on the spatial twist $mg+nh$ and the temporal twist $m'g+n'h$.

To confirm that the one given in \eqref{foo} is a correct choice, 
we only have to take a typical example,
using the universality of the anomaly-cancelling phases.
So, let the original theory $T=T(\Lambda)$ be a chiral boson theory 
on an even-self-dual lattice $\Lambda\subset \bR^n$,
which has a $U(1)^n$ symmetry parameterized by $\bR^n/\Lambda$.
We take a subgroup $\bZ_a\times \bZ_b \simeq \langle g,h\rangle \subset U(1)^n$ chosen as above.

To see when $G$ is anomaly-free, let us try to perform the orbifold stepwise, 
first by $\bZ_a$ and then by $\bZ_b$.
The anomaly-free condition associated to the first step is that \begin{equation}
\frac12\vec\mug\cdot\vec\mug \in  \frac{1}{a}\bZ,
\end{equation}
see  e.g.~\cite{Chang:2018iay}.
Then the generator of $\bZ_a$ acts by $e^{2\pi i \vec\lambda\cdot\vec\mug}$ on 
vertex operators associated to $\vec\lambda\in \Lambda$,
while it acts by $e^{\pi i m\vec\mug\cdot \vec\mug}$ on the  $m\vec\mug$-twisted sector.
One can then check that 
the lattice $\Lambda|_{\vec\mug} \subset \bR^n$ defined by \begin{equation}
\Lambda|_{\vec\mug}:=\{\vec\lambda + m \vec\mug \mid
\vec\lambda\in\Lambda,\ m\in\bZ,\ 
(\vec\lambda+ \frac m2\vec\mug )\cdot\vec\mug\in\bZ\}
\end{equation}
is even and self-dual,
and $T(\Lambda)/\bZ_a = T(\Lambda|_{\vec\mug})$.

To be able to orbifold this $\bZ_a$-orbifolded theory by $\bZ_b$ generated by $\vec\muh$,
we need two conditions, namely 
\begin{equation}
\vec\mug\cdot\vec\muh \in \frac1b \bZ,\quad\text{and}\quad
\frac12\vec\muh\cdot\vec\muh \in  \frac 1b\bZ.
\end{equation}
The first is simply that $\vec\muh$ still generates a $\bZ_b$-symmetry
of $T(\Lambda|_{\vec \mug})$,
and the second is the standard anomaly-free condition for the $\bZ_b$ orbifold.\footnote{%
When $a$ is not chosen to divide $b$, 
the first condition is weakened to $\vec\mug\cdot\vec\muh \in (1/\mathrm{lcm}(a,b))\bZ$.
This is because we can always find a suitable $\vec \lambda\in \Lambda$ such that, 
with $\vec\muh' := \vec \muh + \vec \lambda$,
we have $\vec\mug\cdot\vec\muh' \in (1/b)\bZ$.
Then $\vec\muh'$ generates a $\bZ_b$ symmetry in the chiral boson theory on  $\Lambda|_{\vec\mug}$, 
and we can perform the $\bZ_b$ orbifold with $\vec\muh'$.
Note that this generalization modifies \eqref{foo}.
}
Then the successive orbifold $T/\bZ_a/\bZ_b=T(\Lambda|_{\vec\mug})/\bZ_b$ is 
the theory  $T(\Lambda|_{\vec\mug}|_{\vec\muh})$. 
It is straightforward to check that the partition function of this orbifold is exactly the one given by \eqref{foo}.
Note that the term $2\pi i m'n\vec\mug\cdot\vec\muh$ is missing in the exponent.
This is due to our sequential construction of this particular $\bZ_a\times \bZ_b$ orbifold.

We note that, once \emph{one} consistent choice of the anomaly-canceling phase for the subgroup $\langle g,h\rangle$ is found,
all other consistent choices should be obtained by further stacking 
$\langle g,h\rangle$-symmetric invertible phases.
As we are dealing with fermionic theories,
possible invertible phases are either (i) the Arf theory itself, 
(ii) $q(f)$ where $f:\bZ_a\times \bZ_b \to \bZ_2$ and $q$ is a quadratic refinement associated to the spin structure,
or (iii) standard bosonic discrete torsion phases for $\bZ_{\mathrm{gcd}(a,b)}=\bZ_a$.\footnote{%
For example, when $a=b$, we can replace the role of $g$ and $h$
in \eqref{foo}.
The phase in \eqref{foo} is not symmetric under this exchange,
but the two choices are different by the standard $\bZ_a\times \bZ_a$ discrete torsion
$e^{2\pi i(mn'-m'n)\vec \mug \cdot \vec \muh}$, as dictated by the general theory.
}

Coming back to our Conway symmetric theories, 
we have  $n=24\times 12$, and the vectors $\vec\mug,\vec\muh\in \bR^{24\times 12}$
are obtained by taking 24 copies of $\vec\lambda,\vec\mu\in \bR^{12}$
associated to the commuting pair $(g,h)$.
In total, our condition is that \begin{equation}
\alpha(g,h)=\pm e^{2\pi i 24\vec\lambda\cdot\vec\mu} e^{2\pi i k/a}
\label{foofinal}
\end{equation} for some choice of the sign $\pm$ and the number $k\in \bZ_a$.

Note that once the anomaly-canceling phase is chosen for the subgroup $\langle g,h\rangle$,
the phases of the twisted partition functions for all other $(g^a h^b, g^c h^d)$ 
are also fixed. 
This information in principle can be used to generate more consistency conditions,
but we do not need and did not use this additional information in our analysis.

\subsection{Partition functions of permutation orbifolds}

There is one final ingredient we need to explain before we can compute the partition functions of the orbifolds we are interested in.
Namely, we need the partition function of the permutation orbifold $T^n/S_n$
and the alternating orbifold $T^n/A_n$, 
starting from the partition function of a given theory $T$.
Although there is the possibility of including the $\bZ_2$ discrete torsion phase in the $S_n$ or $A_n$ orbifold,
we will not do so in this paper.

When the $Spin(24)_{n}$ symmetry is ignored, the way to compute
the partition function of the permutation orbifold $T^n/S_n$ 
is well-established, together with 
its generalization to the alternating orbifold $T^n/A_n$.
For details, see \cite{Albert:2022gcs} and the references therein.
What was left unfinished was to properly incorporate the $Spin(24)_{n}$ symmetry into the computation.

For this purpose, it is convenient to consider the generating function of the partition functions 
for all $n$:
\begin{align}
\cZ[T]^{\alpha \vec\mu}_{\beta\vec\lambda}(\tau)&:=
\sum_{n=0}^\infty p^n Z[T^n/S_n]^{\alpha \vec\mu}_{\beta\vec\lambda}(\tau)\\
&=
\exp\Bigl(\sum_{n=1}^\infty p^n \sT_nZ[T]^{\alpha \vec\mu}_{\beta\vec\lambda}(\tau)\Bigr).
\end{align}
Here, the parameters $\alpha,\beta\in \{0,1\}$ specify the 
holonomies of $\bZ_2=S_n/A_n$ in the temporal and spatial directions, 
whereas the parameters $\vec\mu,\vec\lambda\in \bR^{24}$ specify the holonomies of $Spin(24)$ in the temporal and spatial directions,
respectively.
Then, we have\begin{align}
Z[T^n/S_n]^{\vec\mu}_{\vec\lambda}(\tau)&=Z[T^n/S_n]^{0\vec\mu}_{0\vec\lambda}(\tau),\\
Z[T^n/A_n]^{\vec\mu}_{\vec\lambda}(\tau)&=\frac12\sum_{\alpha,\beta\in \{0,1\}} Z[T^n/S_n]^{\alpha\vec\mu}_{\beta\vec\lambda}(\tau).
\end{align}

We now need to specify the action of the Hecke operator $\sT_n$:
\begin{align}
    \sT_nZ[T]^{\alpha \vec\mu}_{\beta\vec\lambda}(\tau) &:= 
    \frac1n\sum_{a d= n } \sum_{0\le b < d} 
    (-1)^{\alpha a(d+1)}(-1)^{\beta ((a+1)d + b(d+1))} \\
&\qquad\times    e^{-\pi i d b \vec\lambda\cdot \vec\lambda }
    Z[T]^{a\vec\mu-b\vec\lambda}_{d\vec\lambda}(\frac{a\tau + b}{d}).
\end{align}
The part independent of the $Spin(24)$ holonomies is the same as in \cite{Albert:2022gcs}, to which we refer the reader  for details.
To explain the dependence on $\vec \mu$ and $\vec \lambda$,
let us recall that this Hecke formula
arises by considering various $n$-fold covers of the torus with a given modular parameter $\tau$,
which are then sewn together using spatial and horizontal $S_n$ holonomies.
An $n$-fold cover splits into a number of connected components,
each of which is a torus whose three vertices are at $0$, $d$ and $a\tau+b$,
for some $a,b,d$ satisfying $ad=n$ and $0\le b<d$.
This new torus has the modular parameter $\frac{a\tau+b}{d}$.

Now, on the original torus, we have  holonomies $\vec\mu$ and $\vec\lambda$ in the temporal and spatial directions, respectively.
On the connected component of the cover with three vertices at $0$, $d$ and $a\tau+b$,
the holonomies in the temporal and spatial directions are $a\vec\mu-b\vec\lambda$ and $d\vec\lambda$, respectively,
where the minus sign in $-b\vec\lambda$ is due to our convention,
see Sec.~\ref{sec:theta-conventions}.
What remains to be explained is the appearance of the phase $e^{-\pi db i \vec\lambda\cdot \vec\lambda}$ in the formula.

\begin{figure}
\centering
\begin{tikzpicture}

\begin{scope}
    \draw[orange!80!yellow, very thick] (0,0.5) -- node[pos=1/3, above, black, font=\normalsize] 
        {$a \vec \mu$} (6,0.5);

    \draw[cyan, very thick]
        (1,0) -- (1,1)
        arc[start angle=180, end angle=90, radius=0.5]
        -- (2.5,1.5)
        arc[start angle=-90, end angle=0, radius=0.5]
        -- node[midway, right, black, font=\normalsize] 
        {$\vec\lambda$} (3,3)
        arc[start angle=180, end angle=90, radius=0.5]
        -- (4.5,3.5)
        arc[start angle=-90, end angle=0, radius=0.5];

    \draw[cyan, very thick]
        (3,0) -- (3,1)
        arc[start angle=180, end angle=90, radius=0.5]
        -- (4.5,1.5)
        arc[start angle=-90, end angle=0, radius=0.5]
        -- node[midway, right, black, font=\normalsize] 
        {$\vec\lambda$} (5,3)
        arc[start angle=180, end angle=90, radius=0.5]
        -- (6,3.5);

    \draw[cyan, very thick]
        (5,0) -- (5,1)
        arc[start angle=180, end angle=90, radius=0.5]
        -- (6,1.5);

    \draw[cyan, very thick]
        (0,1.5) -- (0.5,1.5)
        arc[start angle=-90, end angle=0, radius=0.5]
        -- node[midway, right, black, font=\normalsize] 
        {$\vec\lambda$} (1,3)
        arc[start angle=180, end angle=90, radius=0.5]
        -- (2.5,3.5)
        arc[start angle=-90, end angle=0, radius=0.5];

    \draw[cyan, very thick]
        (0,3.5) -- (0.5,3.5)
        arc[start angle=-90, end angle=0, radius=0.5];

    \draw[dashed, thick] (1,1.5) circle (0.55);

    \draw[thick] (0,0) rectangle (6,4);
\end{scope}

\draw[->, thick] (6.5,2) -- (7.4,2);

\begin{scope}[xshift=8cm]
    \draw[orange!80!yellow, very thick] (0,0.5) -- (6,0.5);

    \draw[cyan, very thick]
        (1,0) -- (1,4);

    \draw[cyan, very thick]
        (3,0) -- (3,4);

    \draw[cyan, very thick]
        (5,0) -- (5,4);

    \draw[cyan, very thick]
        (0,1.5) -- (6,1.5);

    \draw[cyan, very thick]
        (0,3.5) -- (6,3.5);

    \draw[dashed, thick] (1,1.5) circle (0.55);

    \draw[thick] (0,0) rectangle (6,4);
\end{scope}

\end{tikzpicture}
\caption{$bd$ manipulations are need to straighten the $d$ topological lines
labeled by $\lambda$.
The figure is for the case $d=3$ and $b=2$.
}
\label{fig:manipulation}
\end{figure}
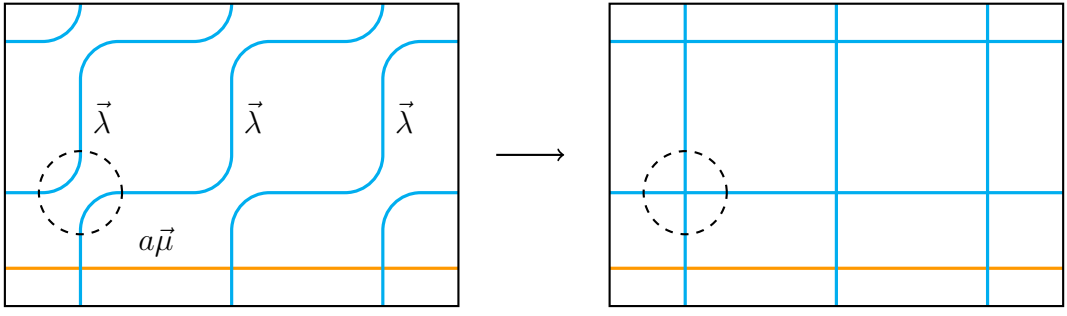

For this purpose, it is convenient to represent the holonomies
by topological lines. On the original torus, we have 
one vertical topological line labeled by $\vec\lambda$.
In the connected component of the cover with three vertices at $0$, $d$ and $a\tau+b$,
we then have $d$ vertical topological lines labeled by $\vec\lambda$,
but if we regard this component as a torus with modular parameter $\frac{a\tau+b}{d}$,
the $d$ vertical lines go in an oblique direction.
We need $bd$ topological manipulations to straighten them up,
as shown in Fig.~\ref{fig:manipulation}.

The phase associated to each of these manipulations should be computable
from first principles, if we set up our conventions for the torus partition functions and the topological lines carefully enough.
It is easier, however, to determine it by considering the case 
when $d=1$ and $b=1$, which is simply the $T$ transformation of the torus partition function.
Referring to the conventions set up in Sec.~\ref{sec:theta-conventions},
we find that this phase is indeed $e^{-\pi i \vec\lambda\cdot \vec\lambda}$.\footnote{
Again, this discussion shows clearly that the phase $e^{-\pi i db \vec\lambda\cdot\vec\lambda}$ appearing here depends on
the conventions, 
and cannot be applied to other conventions without modification,
so the readers should be careful about this point.
}

\subsection{Final results for the orbifold partition functions}
\label{sec:PFalgorithm}

We can now combine all of the information described so far
to compute the partition functions of the orbifolds we are interested in.
We use the two theories $T=V^{24}/A_{24}$ and $\tilde T=\tilde V^{24}/A_{24}$,
the alternating orbifolds for $V$ and $\tilde V=V\times \Arf$,
to impose the integrality conditions \eqref{ABeq} for the anomaly-cancelling phases $A_\cO$ and $B_\cO$.
As the possibilities for each $A_\cO$ and $B_\cO$ are finite,
we can in principle determine all of them by brute force.
But we cannot be too naive. For example, enumerating all possibilities
is impossible. Even $A_\cO=\pm1$ alone gives $2^{639}$ possibilities, which is too large to enumerate.
It was not immediately clear to the authors how best to proceed.
One method that worked, although clearly not the most efficient one,
was as follows.

Each of the equation \eqref{ABeq}, with $T$, $g$ and $\chi$ fixed,
is fairly sparse, and  the coefficients multiplying $A_\cO$ and $B_\cO$
are often zero. There are a few reasons for this.
First, only those $SL(2,\bZ)$ orbits $\cO$ which contain commuting pairs of the form $(g,h)$ for some $h$ contribute.
Second, the coefficients involve the characters $\chi(h)$, 
which are often zero.
Third, $Z[T]_g^h$ is also often zero, from various reasons.

Using this observation, we can set up the following algorithm.
In the following, $U$ is the set of $SL(2,\bZ)$ orbits $\cO$ for which $A_\cO$ and $B_\cO$ are not yet determined. 
Originally, it contains all $SL(2,\bZ)$ orbits.
We gradually remove orbits from $U$ as we determine the values of $A_\cO$ and $B_\cO$, and aim to shrink $U$ to the empty set.

Fix a theory $T$ with $\Co_0$ symmetry. Then, for a fixed theory $T$,
we perform the following loop:
\begin{enumerate}
\item Look for $g$ and $\chi$ such that the equation \eqref{ABeq} contains only
 a fairly small number, say $10$, of $A_\cO$ and $B_\cO$ with $\cO\in U$.
\item Fix such $g$ and $\chi$, and let $S\subset U$ be the subset of orbits 
appearing in \eqref{ABeq} for this choice of $g$ and $\chi$.
\item Let $\Pi$ be the set of all possibilities for the $A_\cO$ and $B_\cO$ in $S$.
We use the notation $A_\cO^p$ and $B_\cO^p$ for the corresponding values of $A_\cO$ and $B_\cO$ for each choice $p\in \Pi$.
Evaluating \eqref{ABeq} for each $p$, determine the subset $\Pi'\subset \Pi$ of possibilities satisfying the integrality condition \eqref{ABeq}.
\item Go over each $\cO\in S$ to see if $A_\cO^p$ (or $B_\cO^p$) is independent of $p$ for all $p\in \Pi'$.
If so, $A_\cO$ (or $B_\cO$) is determined to take this value common for all $p\in \Pi'$,
and we can remove $\cO$ from $U$.
\item Go back to step 1, until all $(g,\chi)$ giving nontrivial constraints
on $A_\cO$ and $B_\cO$ for $\cO\in U$ are exhausted.
\end{enumerate}

To speed up the process, we have also used the following additional information,
applicable only to our particular case, namely that \begin{equation}
\hat Z[T]_e^h = Z[T]_e^h
\end{equation}for all conjugacy classes $[h]$ of $\Co_0$, where $e$ is the identity element.
To see this, we again use the universality of the anomaly-cancelling phases,
and study the case when $T=V^{24}$.
Then, the action of $\Co_0\subset Spin(24)$ 
is uniquely fixed on the untwisted sectors of $T$
from the Hamiltonian interpretation,
and by inspecting the explicit formulas for $Z[T]_e^h$
which we set up in Sec.~\ref{sec:theta-conventions},
we see that the anomaly-cancelling phases are trivial for these cases.
This allows us to set $A_\cO$ and $B_\cO$ for the orbits $\cO$ containing 
commuting pairs of the form $(e,h)$ for some $h$.
This happens to fix 160 $A_\cO$'s out of 639 $A_\cO$'s in total.
We then start the algorithm described above.

Now, there is no \emph{a priori} guarantee that this procedure would terminate
with a unique solution for all $A_\cO$ and $B_\cO$.
But in practice, by applying it to $T=V^{24}/A_{24}$,
the size of $U$ already shrunk to 27.
The phases for the remaining orbits were then fixed  by applying 
the same algorithm to $\tilde T=\tilde V^{24}/A_{24}$.
We implemented this algorithm in Mathematica, as detailed in Appendix~\ref{app:PF}.\footnote{%
For those readers who do not have licenses, we created a video of how it runs
and posted at YouTube at \url{https://www.youtube.com/watch?v=5qP8mt01PjM}.
It takes about 10 minutes on an M1 Mac laptop.
}

Using the determined anomaly-cancelling phases $A_\cO$ and $B_\cO$,
we can finally compute the partition functions of the orbifolds we are interested in.
We find that\begin{equation}
Z[V^{24}/\Co_0]=1331666454021264\sim 1.33\times 10^{15},
\end{equation}
from which we infer that \begin{equation}
Z[V^{24}/\Co_1]=665833227010632\sim 6.66\times 10^{14}.
\end{equation}
We also find that the partition functions of the permutation orbifolds are
 \begin{align}
Z[V^{24}/S_{24}\times\Co_0]&=81270,&
Z[V^{24}/A_{24}\times\Co_0]&=162540,\\
Z[\tilde V^{24}/S_{24}\times\Co_0]&=122374,&
Z[\tilde V^{24}/A_{24}\times\Co_0]&=244748,
\end{align}
where we remind the reader that $\tilde V=V\times \Arf$.
We see that the values of the partition functions 
for the alternating orbifolds are exactly twice those of the symmetric orbifolds,
which is consistent with our analysis in Sec.~\ref{sec:our-orbifolds}.
We note that the same anomaly-cancelling phases can be used to compute the torus partition functions with other spin structures and read off the full spectrum of these theories,
but we do not report them here.

After all these computations, against the authors' original hope,
it turned out that these permutation-Conway orbifolds
do not provide us the elusive periodicity element of
topological modular forms, which should have $Z=1$.
Still, the obtained partition function attains 
the smallest known value for
the partition function of a Conway symmetric theory
with this chiral central charge and with a single NS-sector vacuum.
Also, we have learned a lot about the structure of the Conway group,
how to manipulate the commuting pairs,
and the way the anomaly-cancelling phases can be determined.
Hopefully, these techniques would be useful in the future,
to the authors and to the readers, to perform orbifold computations
in other cases where the symmetry group is huge and
the anomaly-cancelling phases are nontrivial.
For example, we have not computed the partition functions
when we include $\bZ_2$ discrete torsion phases in the $S_n$ or $A_n$ orbifold.
We might also entertain the orbifold with respect to other nontrivial subgroups of the form $G\subset S_{24}$,
possibly the largest Mathieu group.
The authors hope to come back to these issues in the future.

\newpage
\section*{Acknowledgments}
The authors thank Ying-Hsuan Lin and Justin Kaidi for their collaboration in the early part of the project,
and illuminating discussions throughout the entirety of the project.
The authors would like to thank them also for carefully proofreading the almost final draft,
and suggesting  many points that could be improved.
The authors also thank Gerald H\"ohn for the information that the Conway group can be constructed
with the standard computer algebra system for group theory (GAP, Magma) with the memory available
for personal computers in 2025.
The authors would also like to thank Yasunori Lee and Yi Zhang for useful discussions.

YT is supported in part by WPI Initiative, MEXT, Japan at Kavli IPMU, the University of Tokyo
and by JSPS KAKENHI Grant-in-Aid (Kiban-C), No.24K06883.
MO was also supported in part by WPI Initiative, MEXT, Japan at Kavli IPMU, the University of Tokyo.
While affiliated with the university, MO was supported by FoPM (WINGS Program of the University of Tokyo), JSPS Research Fellowship for Young Scientists, and
JSPS KAKENHI Grant No.\ JP23KJ0650.
JA acknowledges partial support from Simons Foundation grant 917464 (Simons Collaboration on Confinement and QCD Strings). Part of the numerical computations in this work were performed on the IPMU gw cluster and
the Princeton Della cluster.

\section*{Statements on the AI usage}

The authors did not use the research-level AI tools 
(which became available after around early summer 2026) at all,
as the research was completed around late April of 2026.
The writing of the paper was delayed because the authors had other obligations to tend to.
The only AI use in this paper was the use of Github Copilot 
to provide auto-complete while writing the LaTeX files.

With the help of the latest research-levelAI tools, 
the project could have been completed in a much shorter time.
For example, 
at first, the authors did not know 
how to enumerate and manipulate the commuting pairs of the Conway group.
The authors had to email a number of group theorists who
they did not know personally to ask for help,
but nobody replied.

It took the authors a few months to recall
that Gerald H\"ohn,
with whom the authors had some prior contacts,
was also an expert on sporadic groups.
When emailed, he was kind enough to immediately reply to the authors with the 
information that it could be done on a personal computer
using GAP or Magma.
With the modern research-level AI tools, this process
would have only taken a few minutes rather than a few months, 
as the AI tools know all the relevant mathematical literature.
The same can be said about the coding parts. 

Writing up this paper in autumn of 2026 made the authors 
nostalgic about the olden days.
It would have been fun to see if the AI tools 
can work out the content of this paper independently, all by themselves.
But after putting this paper on the arXiv,
there would be no way to see this,
as the AI tools would read this paper and learn from this immediately.

\appendix
\section{Computational details}\label{app:ComputDetails}
Here we provide more details of our computations,
which used GAP \cite{GAP4}, C++ and Mathematica.\footnote{%
Recently it was renamed to the Wolfram language, 
but the authors prefer to use the old name,
just as they prefer to call a certain social network service using its old ornithological name rather than a single alphabet.
In any case, now that SymPy and related softwares have become 
more and more powerful, it would be preferable to use these open softwares
rather than the proprietary language.
Unfortunately the authors did not have the time to do so for this paper,
although it would not have been too difficult to do so
using the modern research-level AI tools.
}
The programs used for the computations and the data produced are all available 
at the third author's GitHub repository 
at \url{https://github.com/yujitach/conway-orbifold/}.

\subsection{Group theory}
\label{app:GAP}

This part of the computation is kept under the subdirectory \texttt{group-theory/} of the repository.
It is implemented in GAP \cite{GAP4} and Mathematica.
\subsubsection{Conway groups as permutation groups}
\label{app:Co}
Let us first explain how we work with $\Co_0$ and $\Co_1$ in practice.
We need to create them as permutation representations.
An explicit description can be found e.g.~in 
a MathOverflow article \cite{MathOverflowArticle}
or in \cite[Chapter 11]{CurtisBook}.
As we did not know either of the references, we proceeded as follows.

According to \cite{ATLASweb},
the Conway group $\Co_0$ is generated by two integral $24\times 24$ matrices $g_1$ and $g_2$, given as follows:

{\tiny
\begin{equation}
\arraycolsep=.5\arraycolsep
A=\left(
\begin{array}{cccccccccccccccccccccccc}
 2 & 0 & 0 & -3 & -2 & -1 & 0 & -2 & 1 & 1 & 0 & 1 & 1 & 0 & 0 & 1 & -1 & -1 & 0 & 0 & -1
   & 0 & 0 & 0 \\
 -4 & 1 & 0 & 6 & 4 & 2 & 2 & 3 & -1 & -3 & -1 & 0 & -2 & -1 & -1 & -1 & 1 & 1 & 0 & 1 &
   0 & 0 & 0 & 0 \\
 4 & -1 & 0 & -6 & -4 & -2 & -3 & -4 & 2 & 4 & 3 & -3 & 2 & 1 & 2 & 0 & 1 & -1 & -1 & -1
   & 1 & 0 & -1 & 1 \\
 4 & 1 & 1 & -6 & -3 & -2 & -1 & -4 & 2 & 2 & 1 & 0 & 1 & 0 & 1 & 2 & 0 & -1 & -1 & 0 &
   -1 & 0 & 0 & 1 \\
 5 & -2 & 0 & -8 & -5 & -3 & -2 & -4 & 2 & 4 & 2 & 0 & 2 & 2 & 1 & 1 & -1 & -1 & 0 & 0 &
   0 & 1 & -1 & 0 \\
 -3 & 0 & -1 & 4 & 1 & 2 & 1 & 1 & 0 & -2 & 0 & 0 & 0 & -1 & -1 & -1 & 0 & 0 & 0 & 0 & -1
   & -1 & 0 & 0 \\
 1 & 0 & -1 & -2 & -2 & -1 & 0 & -2 & 0 & 1 & 1 & 0 & 0 & 0 & 0 & 0 & -1 & 0 & 0 & -1 & 0
   & 0 & 0 & 0 \\
 -3 & 1 & -1 & 5 & 3 & 1 & 2 & 2 & -1 & -2 & -1 & 0 & -1 & -1 & -1 & -1 & 0 & 0 & 1 & 0 &
   0 & 0 & 0 & 0 \\
 -2 & 2 & 0 & 3 & 3 & 0 & 2 & 2 & 0 & -1 & -2 & 1 & 0 & -1 & 0 & 0 & 1 & -1 & 1 & 1 & 0 &
   0 & 0 & 1 \\
 9 & -1 & -1 & -13 & -9 & -5 & -4 & -9 & 4 & 6 & 4 & -1 & 3 & 1 & 2 & 2 & -1 & -3 & -1 &
   -1 & -2 & 1 & -1 & 1 \\
 1 & -1 & -1 & -3 & -3 & -1 & -1 & -4 & 2 & 2 & 3 & -2 & 1 & 0 & 0 & 0 & 0 & 0 & -1 & -1
   & 0 & 0 & -1 & 1 \\
 4 & -1 & 1 & -6 & -4 & -2 & -2 & -3 & 2 & 3 & 2 & -1 & 2 & 1 & 1 & 1 & 0 & -1 & -1 & 0 &
   0 & 0 & -1 & 1 \\
 -4 & 0 & 0 & 5 & 3 & 2 & 2 & 3 & -2 & -2 & -1 & 0 & -2 & 0 & -1 & -1 & 0 & 2 & 0 & 0 & 1
   & 0 & 0 & -1 \\
 8 & 2 & 2 & -9 & -5 & -4 & 0 & -5 & 2 & 2 & 2 & 1 & 0 & 1 & 1 & 2 & -1 & 0 & -1 & 1 & -1
   & 1 & 0 & 1 \\
 -3 & 0 & 0 & 3 & 3 & 1 & 0 & 3 & -1 & -1 & -1 & 0 & 0 & 0 & 0 & 0 & 1 & 0 & 1 & 0 & 1 &
   -1 & 0 & 0 \\
 -5 & 2 & 0 & 6 & 4 & 3 & 3 & 2 & -1 & -4 & -2 & 1 & -2 & -2 & -2 & 0 & 0 & 1 & 0 & 0 &
   -1 & -1 & 1 & 0 \\
 3 & 0 & 0 & -3 & -3 & -1 & 0 & -2 & 0 & 1 & 1 & 0 & 0 & 0 & 0 & 0 & -1 & 0 & -1 & -1 & 0
   & 0 & 0 & 0 \\
 0 & 0 & 0 & 1 & 1 & 0 & 0 & 2 & -1 & 0 & -2 & 1 & 0 & 0 & 0 & 0 & 0 & -1 & 1 & 0 & 0 & 0
   & 0 & -1 \\
 -1 & 0 & -1 & 2 & 0 & 1 & -1 & 0 & 0 & 0 & 0 & -1 & 1 & -1 & 0 & -1 & 1 & -1 & 0 & -1 &
   0 & -1 & 0 & 0 \\
 -4 & -1 & -1 & 4 & 2 & 2 & 0 & 2 & -1 & -2 & -1 & 1 & -1 & -1 & -1 & 0 & 0 & 1 & 0 & 0 &
   -1 & 0 & 1 & -1 \\
 -9 & 1 & 0 & 11 & 7 & 5 & 3 & 6 & -2 & -5 & -3 & 1 & -2 & -2 & -2 & -1 & 1 & 2 & 0 & 1 &
   0 & -1 & 1 & 0 \\
 -6 & 1 & 0 & 9 & 5 & 4 & 3 & 4 & -2 & -4 & -1 & -1 & -2 & -1 & -1 & -2 & 1 & 2 & 0 & 0 &
   1 & -1 & 0 & 0 \\
 5 & 1 & -1 & -7 & -5 & -3 & -1 & -6 & 2 & 3 & 3 & 0 & 1 & 0 & 1 & 1 & -1 & 0 & -1 & -1 &
   -1 & 1 & 0 & 1 \\
 6 & 0 & 1 & -6 & -4 & -2 & -2 & -3 & 1 & 2 & 1 & 0 & 2 & 1 & 1 & 1 & -1 & -1 & 0 & 0 &
   -1 & 0 & 0 & 0 \\
\end{array}
\right)
\end{equation}}
and {\tiny\begin{equation}
\arraycolsep=.5\arraycolsep
B=\left(
\begin{array}{cccccccccccccccccccccccc}
 0 & 0 & 1 & 0 & 0 & 0 & -1 & 1 & 1 & 0 & -1 & 0 & 1 & 0 & 1 & 0 & 1 & 0 & 0 & 1 & 0 & 1
   & 0 & 1 \\
 3 & 1 & -1 & -3 & -2 & -2 & 0 & -3 & 1 & 2 & 2 & -1 & 1 & 0 & 1 & 0 & 0 & -1 & 0 & -1 &
   0 & -1 & -1 & 0 \\
 -4 & 0 & -1 & 7 & 4 & 3 & 1 & 4 & -3 & -3 & -2 & 0 & -1 & -1 & -1 & -1 & 0 & 0 & 1 & -1
   & 0 & -1 & 1 & -1 \\
 0 & -1 & 0 & 1 & 0 & 1 & -1 & 1 & 0 & 0 & 0 & -1 & 1 & 0 & 1 & -1 & 0 & 0 & 0 & 0 & 0 &
   0 & 0 & 0 \\
 6 & 0 & 2 & -6 & -4 & -2 & -3 & -2 & 1 & 2 & 1 & -1 & 1 & 1 & 2 & 1 & 1 & -1 & -1 & 0 &
   0 & 1 & 0 & 1 \\
 2 & 0 & -1 & -5 & -3 & -3 & -1 & -4 & 3 & 3 & 1 & 0 & 2 & 0 & 1 & 1 & 0 & -1 & 0 & 0 & 0
   & 1 & -1 & 1 \\
 -10 & 0 & 0 & 12 & 8 & 5 & 4 & 7 & -3 & -6 & -3 & 1 & -3 & -1 & -3 & -1 & 0 & 3 & 1 & 1
   & 1 & -1 & 1 & -1 \\
 -1 & 4 & 1 & 5 & 4 & 1 & 3 & 3 & -2 & -3 & -2 & 1 & -1 & -1 & 0 & 0 & 1 & 1 & 1 & 0 & 0
   & -1 & 1 & 0 \\
 -1 & 0 & 1 & 3 & 2 & 1 & -1 & 3 & -1 & -1 & -2 & 0 & 1 & 0 & 1 & 0 & 1 & 0 & 1 & 0 & 0 &
   0 & 1 & 0 \\
 2 & 2 & 1 & 0 & 0 & 0 & 0 & 0 & 0 & 0 & 0 & -1 & 1 & 0 & 2 & 0 & 1 & 0 & 0 & 0 & 0 & 0 &
   0 & 1 \\
 -5 & 3 & 0 & 11 & 7 & 4 & 4 & 6 & -4 & -5 & -3 & 0 & -2 & -2 & -1 & -2 & 1 & 1 & 1 & -1
   & 1 & -2 & 1 & -1 \\
 3 & 0 & 1 & -2 & -1 & -1 & -1 & 0 & 0 & 1 & -1 & 0 & 1 & 0 & 1 & 0 & 0 & -1 & 0 & 0 & 0
   & 1 & 0 & 0 \\
 -4 & 0 & -1 & 5 & 3 & 2 & 2 & 2 & -1 & -3 & 0 & 0 & -2 & -1 & -2 & -1 & 0 & 1 & 0 & 0 &
   0 & -1 & 0 & -1 \\
 2 & -1 & 1 & -2 & -1 & -1 & -1 & 0 & 1 & 1 & 1 & -1 & 1 & 1 & 1 & 0 & 0 & 0 & 0 & 1 & 1
   & 0 & -1 & 0 \\
 2 & -1 & 1 & -3 & -1 & -1 & -1 & 0 & -1 & 1 & -1 & 1 & -1 & 1 & 0 & 1 & -1 & 0 & 0 & 0 &
   0 & 1 & 1 & -1 \\
 -2 & 0 & -1 & 3 & 1 & 1 & 1 & 1 & 0 & -1 & 0 & 0 & 0 & -1 & 0 & -1 & 0 & 1 & 0 & 0 & 0 &
   0 & 0 & 0 \\
 -11 & -2 & -1 & 12 & 7 & 5 & 3 & 7 & -2 & -5 & -3 & 1 & -2 & -1 & -3 & -2 & 0 & 2 & 1 &
   1 & 1 & 0 & 0 & -1 \\
 -2 & 0 & 1 & 2 & 2 & 1 & 0 & 3 & -1 & -1 & -2 & 1 & -1 & 0 & 0 & 0 & 1 & 0 & 0 & 1 & 0 &
   1 & 0 & 0 \\
 -6 & 1 & -1 & 7 & 5 & 2 & 2 & 4 & -2 & -2 & -3 & 1 & -1 & -1 & -1 & -1 & 1 & 0 & 1 & 0 &
   1 & 0 & 0 & 0 \\
 -3 & 1 & -1 & 3 & 2 & 1 & 2 & 0 & 0 & -1 & 0 & 0 & -1 & -1 & -1 & -1 & 0 & 0 & 0 & 0 & 0
   & -1 & 0 & 0 \\
 2 & -2 & -2 & -6 & -5 & -2 & -2 & -6 & 3 & 4 & 3 & -1 & 2 & 0 & 0 & 0 & -1 & -1 & -1 &
   -1 & -1 & 0 & -1 & 0 \\
 2 & -1 & -2 & -4 & -4 & -1 & -1 & -5 & 2 & 2 & 4 & -2 & 1 & 0 & 0 & 0 & -1 & 0 & -1 & -2
   & 0 & -1 & -1 & 0 \\
 -6 & 1 & 0 & 8 & 5 & 4 & 3 & 3 & -2 & -4 & -1 & 0 & -2 & -1 & -2 & -1 & 0 & 2 & 0 & 0 &
   0 & -2 & 1 & -1 \\
 14 & 1 & 2 & -18 & -11 & -7 & -4 & -10 & 4 & 7 & 4 & 0 & 3 & 2 & 3 & 3 & -1 & -2 & -2 &
   0 & -1 & 2 & -1 & 2 \\
\end{array}
\right)
\end{equation}}
They are known to satisfy the two relations\footnote{%
These relations are also taken from the web ATLAS \cite{ATLASweb}.
It turns out that there are version 2 and version 3 webpages of the web ATLAS.
Furthermore, quite confusingly, the version 2 webpage is still maintained, 
while the version 3 webpage is not.
For example, compare the version 2 and version 3 webpages for the Conway group $\Co_1$, which are at
\url{https://brauer.maths.qmul.ac.uk/Atlas/spor/Co1/} and 
\url{https://brauer.maths.qmul.ac.uk/Atlas/v3/spor/Co1/}.
The v3 page has a typo in the second relation, which
lacks the last $B$.
The still-maintained v2 page presents the correct relations.
}
\begin{equation}
B^3=e,\qquad (ABABABABBABABBABB)^{33}=e,
\label{conway-relations}
\end{equation}
where we used $A=g_1$ and $B=g_2$ for brevity.

Unfortunately the inner product on the Leech lattice $L=\bZ^{24}$ is not given there. 
But it can be recovered. 
Let $M$ be a symmetric $24\times 24$ matrix representing the inner product. 
This should be invariant under $g_1$ and $g_2$, i.e.~we demand \begin{equation}
g_1 M g_1^T = M, \qquad
g_2 M g_2^T = M. \label{inv}
\end{equation}
This determines $M$ uniquely up to scalar multiplication. 
Normalizing every matrix entry to be an integer and demanding the GCD to be one, we find that $M$ is uniquely given by 

{\tiny
\begin{equation}
\arraycolsep=.5\arraycolsep
M=\left(
\begin{array}{cccccccccccccccccccccccc}
 4 & -2 & -2 & 2 & 2 & 2 & -1 & -1 & 2 & 2 & -2 & 2 & -2 & 1 & -1 & 1 & 1 & 2 & 0 & -1 &
   0 & -2 & -2 & 2 \\
 -2 & 4 & 0 & 0 & -2 & 0 & -1 & 2 & 0 & 0 & 2 & -1 & 1 & 1 & -1 & 1 & -2 & -2 & -1 & 1 &
   1 & 2 & 1 & -2 \\
 -2 & 0 & 4 & 0 & 0 & -2 & 0 & 0 & 0 & 0 & 2 & 0 & 0 & -2 & 1 & -1 & -1 & -1 & 1 & -1 &
   -1 & 1 & 0 & -1 \\
 2 & 0 & 0 & 4 & 0 & 0 & -2 & -1 & 2 & 2 & 0 & 2 & -2 & 1 & 0 & 2 & -1 & 0 & -1 & 0 & 1 &
   -1 & -1 & 0 \\
 2 & -2 & 0 & 0 & 4 & 0 & -1 & -1 & 1 & 1 & -1 & 1 & -1 & 0 & 0 & -1 & 0 & 2 & 0 & -2 &
   -2 & -1 & -2 & 2 \\
 2 & 0 & -2 & 0 & 0 & 4 & 0 & 0 & 1 & 1 & -1 & 1 & -1 & 1 & -1 & 1 & 1 & 0 & 1 & 0 & 1 &
   0 & -1 & 1 \\
 -1 & -1 & 0 & -2 & -1 & 0 & 4 & 1 & -1 & -1 & 0 & -2 & 2 & 0 & 0 & 0 & 2 & -1 & 0 & 0 &
   -1 & 1 & 2 & -1 \\
 -1 & 2 & 0 & -1 & -1 & 0 & 1 & 4 & 1 & 1 & 2 & -1 & 1 & 0 & -1 & 1 & -1 & -1 & -1 & -1 &
   -1 & 1 & 1 & -1 \\
 2 & 0 & 0 & 2 & 1 & 1 & -1 & 1 & 4 & 2 & 0 & 2 & -2 & 1 & 0 & 1 & -1 & 0 & -1 & -2 & 0 &
   -1 & -1 & 0 \\
 2 & 0 & 0 & 2 & 1 & 1 & -1 & 1 & 2 & 4 & 0 & 1 & -2 & 0 & -1 & 1 & -1 & 1 & 0 & -1 & -1
   & -1 & -1 & 1 \\
 -2 & 2 & 2 & 0 & -1 & -1 & 0 & 2 & 0 & 0 & 4 & 0 & 1 & -1 & 0 & 1 & -1 & -2 & 0 & 0 & 0
   & 2 & 1 & -2 \\
 2 & -1 & 0 & 2 & 1 & 1 & -2 & -1 & 2 & 1 & 0 & 4 & -2 & 0 & 1 & 0 & 0 & 1 & 0 & -1 & 1 &
   -2 & -2 & 1 \\
 -2 & 1 & 0 & -2 & -1 & -1 & 2 & 1 & -2 & -2 & 1 & -2 & 4 & 0 & -1 & 0 & 1 & -1 & -1 & 1
   & 0 & 2 & 2 & -2 \\
 1 & 1 & -2 & 1 & 0 & 1 & 0 & 0 & 1 & 0 & -1 & 0 & 0 & 4 & -1 & 1 & 0 & -1 & -2 & 0 & 0 &
   0 & 0 & 0 \\
 -1 & -1 & 1 & 0 & 0 & -1 & 0 & -1 & 0 & -1 & 0 & 1 & -1 & -1 & 4 & -1 & -1 & 0 & 0 & 0 &
   0 & -1 & 0 & 0 \\
 1 & 1 & -1 & 2 & -1 & 1 & 0 & 1 & 1 & 1 & 1 & 0 & 0 & 1 & -1 & 4 & 0 & -1 & -1 & 0 & 1 &
   1 & 0 & -1 \\
 1 & -2 & -1 & -1 & 0 & 1 & 2 & -1 & -1 & -1 & -1 & 0 & 1 & 0 & -1 & 0 & 4 & 1 & 1 & 0 &
   0 & 0 & 0 & 0 \\
 2 & -2 & -1 & 0 & 2 & 0 & -1 & -1 & 0 & 1 & -2 & 1 & -1 & -1 & 0 & -1 & 1 & 4 & 1 & -1 &
   -1 & -2 & -2 & 2 \\
 0 & -1 & 1 & -1 & 0 & 1 & 0 & -1 & -1 & 0 & 0 & 0 & -1 & -2 & 0 & -1 & 1 & 1 & 4 & 0 & 0
   & 0 & -1 & 1 \\
 -1 & 1 & -1 & 0 & -2 & 0 & 0 & -1 & -2 & -1 & 0 & -1 & 1 & 0 & 0 & 0 & 0 & -1 & 0 & 4 &
   2 & 0 & 2 & -1 \\
 0 & 1 & -1 & 1 & -2 & 1 & -1 & -1 & 0 & -1 & 0 & 1 & 0 & 0 & 0 & 1 & 0 & -1 & 0 & 2 & 4
   & 0 & 1 & -1 \\
 -2 & 2 & 1 & -1 & -1 & 0 & 1 & 1 & -1 & -1 & 2 & -2 & 2 & 0 & -1 & 1 & 0 & -2 & 0 & 0 &
   0 & 4 & 1 & -2 \\
 -2 & 1 & 0 & -1 & -2 & -1 & 2 & 1 & -1 & -1 & 1 & -2 & 2 & 0 & 0 & 0 & 0 & -2 & -1 & 2 &
   1 & 1 & 4 & -2 \\
 2 & -2 & -1 & 0 & 2 & 1 & -1 & -1 & 0 & 1 & -2 & 1 & -2 & 0 & 0 & -1 & 0 & 2 & 1 & -1 &
   -1 & -2 & -2 & 4 \\
\end{array}
\right).
\end{equation}}
Note that the conditions \eqref{inv} mean that the group elements act on row vectors from the right. 
This also matches the GAP convention that the permutations act on points from the right.
We also learn that all the 24 basis vectors \begin{equation}
e_i = ( 0 \cdots 0 \underbrace{1}_{\text{$i$-th position}} 0 \cdots 0)
\end{equation}
have norm $4$, i.e.~are shortest nonzero vectors of the Leech lattice.

For group theory computations, it is convenient to have a permutation representation of the group.
It is known that $\Co_0$ acts faithfully on the set $\sS$ of norm-$4$ vectors
of the Leech lattice, which has $196560$ elements.
We can generate this set $\sS$ by repeatedly applying $g_1$ and $g_2$ to the basis vectors $e_i$,
and then find the permutation representation of $g_1$ and $g_2$ on $\sS$.

\subsubsection{Conjugacy classes and $SL(2,\bZ)$ actions}
GAP or any other computational group theory software\footnote{%
Magma is another standard software. More recently,
there is also a Julia package called Oscar.jl, which is both a wrapper 
and a reimplementation of GAP.
Any of them would be suitable for our purpose.
} can then be used to construct a permutation group
generated by these two permutations.
We can then enumerate the conjugacy classes, 
 find the centralizers of the representatives of the conjugacy classes,
 and compute the conjugacy classes of the centralizers.
More concretely, we pick a representative $g_i$
from each conjugacy class of $\Co_0$,
and then pick a representative $h_{i,j}$ from each conjugacy class of the centralizer of $g_i$.
Then any conjugacy class of commuting pairs
is of the form $[g_i,h_{i,j}]$.
We call this commuting pair the commuting pair $\#(i,j)$ of $\Co_0$.

One point to note is that the algorithm enumerating the conjugacy classes 
are probabilistic, and therefore the order the conjugacy classes appear in the output is not guaranteed to be the same every time the code is run.
To share data among collaborators, this computation has to be done once
and kept fixed.

Once these data are obtained, we can compute the action of 
$S$, $T$ and other actions on the conjugacy classes of commuting pairs of $\Co_0$.
For example, given a conjugacy class $[g_i,h_{i,j}]$ of commuting pairs,
its image under $S$ is $[h_{i,j},g_i^{-1}]$.
We want to identify this with another conjugacy class $[g_{i'},h_{i',j'}]$.
For this, we first find a $k$ such that $g_{i'}=k h_{i,j} k^{-1}$
for some $i'$.
Then, we look for a $k'$ in the centralizer of $g_{i'}$ such that $h_{i',j'}=k' k g_i^{-1} k^{-1} k'^{-1}$, for some $j'$.
This can be easily done in GAP.

Once this was done, we can study the $SL(2,\bZ)$-orbit structure
of the conjugacy classes of commuting pairs of $\Co_0$.
It can be easily done and visualized using Mathematica,
resulting in Fig.~\ref{fig:co0-orbits}.

The same computations can be done for $\Co_1=\Co_0/\bZ_2$ as well.
In this case, we use a permutation representation
on $\sS/\{\pm 1\}$, the set of norm-4 vectors of the Leech lattice 
modulo the sign, which has $98280$ elements.
After that, everything can be done in the same way, resulting in Fig.~\ref{fig:co1-orbits}.

\subsubsection{Character tables}
We also need the character tables of the centralizers of the representatives of the conjugacy classes.
There is a method called the Dixon-Schneider algorithm,
which can be used to compute the character table of a finite group from its permutation representation.\footnote{%
Very roughly speaking, this method is 
based on computing the fusion ring of the conjugacy classes 
of a given group and then diagonalizing the fusion matrices.
}
It is implemented in GAP, and can then be used to compute the necessary character tables within reasonable time and memory consumption
if a powerful machine is used,
except for the centralizers of the representatives
of \emph{four} conjugacy classes, for which 
the centralizers are too large to be handled 
on a machine with 128GB of memory.

Luckily, the centralizers of the representative of these four conjugacy classes
have a known structure, as can be found by looking up the ATLAS \cite{ATLAS}.
Namely, they are either $\Co_0$ itself,
or $6.\Suz$, the central extension \begin{equation}
1\to \bZ_6\to 6.\Suz\to \Suz\to 1
\end{equation} of another sporadic finite simple group $\Suz$
called the Suzuki group.
For both, the character tables are already known,
given in \cite{ATLAS}, and already digitized and included in GAP.
One issue is that the order of the conjugacy classes 
in the pre-computed character tables 
is not guaranteed to be the same as the order of the conjugacy classes
in the permutation representation we constructed.

To match the order of the conjugacy classes,
we build what can be called a \emph{signature} of each conjugacy class $[g]$,
which consists of the following data:\begin{equation}
\text{signature}([g]) = (\text{order of $g$},
\text{size of $[g]$},
\tr_V g, \tr_V g^2, \ldots, \tr_V g^N),
\end{equation}
where $V$ is a fixed representation of $G$ and $N$ is a fixed positive integer.
The signature assigns an invariant to each conjugacy class,
allowing us to identify the conjugacy classes uniquely in most cases.

This method does not always work, when the group $G$ in question
has \emph{table automorphisms}.
To explain this concept, let $\mathrm{Cl}(G)$ be the set of conjugacy classes of $G$,
and $\sigma_p: \mathrm{Cl}(G)\to \mathrm{Cl}(G)$ be the map sending a conjugacy class $[g]$ to $[g^p]$ for a prime $p$.
A table automorphism is a permutation of $\mathrm{Cl}(G)$ 
that commutes with all $\sigma_p$
and maps a character $\chi$ to another character $\chi'$ of $G$.
For example, when $G$ has an outer automorphism, 
it clearly induces a table automorphism.
But there are table automorphisms that are not induced by outer automorphisms,
and $\Co_0$ and $6.\Suz$ are examples of such groups.

For our purposes, we only need to identify the conjugacy classes 
up to table automorphisms, and the signature method works well enough.
In our case, it sufficed to take $V$ to be the $24$-dimensional representation of $\Co_0$ on the Leech lattice and to take $N=6$.
This method was also needed to assign the standard ATLAS names
to the conjugacy classes of $\Co_0$ we constructed 
using the permutation representation on $\sS$.
The ATLAS names of the conjugacy classes of $\Co_0$ in our ordering are 
as follows: \begin{quotation}
1a, 2a, 6d, 3d, 14b, 7b, 21c, 42c, 23a, 46a, 23b, 46b, 2b, 2c, 6b, 3b, 4b, 4c, 6k, 6l, 8c, 12j, 12i, 24h, 3a, 6a, 6f, 6e, 12d, 12c, 4d, 8e, 8d, 12e, 24d, 24e, 4g, 4a, 8a, 12a, 24a, 12f, 7a, 14a, 28a, 56a, 56b, 8b, 24c, 24f, 12k, 2d, 6p, 4h, 12s, 12b, 8i, 24g, 10b, 5b, 20b, 10j, 6o, 3c, 6c, 4f, 4e, 8g, 8h, 6j, 6i, 12m, 12l, 24j, 24i, 12n, 12o, 6n, 6m, 12r, 12q, 12h, 12g, 10a, 5a, 10e, 10d, 30a, 15a, 30g, 30f, 20d, 20c, 60c, 60d, 26a, 13a, 78a, 39a, 39b, 78b, 8f, 16c, 16b, 6h, 6g, 10f, 10g, 15d, 30d, 30h, 30i, 18a, 9a, 18d, 18e, 36a, 36b, 10c, 5c, 30e, 15e, 10i, 10h, 30j, 30k, 20a, 15c, 30c, 60b, 12p, 20e, 21a, 42a, 84a, 35a, 70a, 40a, 40b, 18b, 9b, 18f, 18g, 9c, 18c, 18i, 18h, 24b, 21b, 42b, 16a, 20g, 20f, 14c, 14d, 52a, 15b, 30b, 60a, 22a, 11a, 22b, 22c, 33a, 66a, 28c, 28b.
\end{quotation}

The same computations can be done for $\Co_1$.
The authors found again that, for two conjugacy classes for which the centralizers
of the representatives are too large,
the Dixon-Schneider algorithm did not finish within reasonable time and memory consumption.
In these cases, the centralizers are $\Co_1$ and $3.\Suz$,
for which the pre-computed character tables were available.
The signature method was then used to identify the conjugacy classes
up to table automorphisms.
The ATLAS names of the conjugacy classes of $\Co_1$ in our ordering are
as follows: \begin{quotation}
1a, 2b, 13a, 26a, 3a, 39a, 39b, 2a, 3b, 4a, 6e, 
8b, 12e, 24e, 2c, 5b, 10f, 10d, 5a, 10a, 5c, 
10e, 3d, 6h, 10b, 15c, 30c, 3c, 6d, 6f, 6c, 
15d, 4c, 12h, 8e, 9c, 7b, 21c, 23a, 23b, 20c, 
7a, 35a, 18c, 21a, 14a, 6b, 4d, 8f, 6a, 12c, 
24d, 4e, 12l, 14b, 28a, 8a, 12a, 24a, 6g, 24c, 
4b, 12g, 12k, 9b, 9a, 4f, 6i, 12m, 11a, 33a, 
8c, 12b, 24b, 8d, 16a, 20a, 15a, 30a, 60a, 15e, 
30e, 21b, 10c, 15b, 30b, 20b, 16b, 22a, 12i, 12d, 
40a, 24f, 30d, 12j, 18b, 42a, 28b, 18a, 36a, 12f.
\end{quotation}

\subsection{Working with the supercurrent}
\label{app:spinor}

We now describe how to compute the partition functions 
of the Duncan theory on the torus with the holonomy of $\Co_0$
specified by a commuting pair $(g,h)$.
We have already set up various conventions of the partition functions
and theta functions in Sec.~\ref{sec:theta-conventions},
and the only remaining issue 
is to bring $g$ and $h$ within the same Cartan torus of $Spin(24)$,
obtaining the vectors $\vec \lambda_g, \vec \lambda_h\in \bR^{12}$.
In this subsection we describe how this can be done in practice.

This part of the computation is kept under the subdirectory \texttt{supercurrent/} of the repository.
It is implemented in Mathematica and C++,
where C++ was used to speed up parts of the computations 
which use $2^{12}\times 2^{12}$ matrices.

\subsubsection{Action on the spinor representation}
Let us first describe how to lift a matrix $g\in SO(24)$
to a matrix $\hat g\in Spin(24)$
as a $2^{12}\times 2^{12}$ matrix acting on the spinor representation.
The same issue at the Lie algebra level is standard in
the high energy physics literature,
and the group-level version is only slightly more complicated.

We first need the Gamma matrices $\Gamma_{i=1,\ldots,24}$.
We construct them using the Jordan-Wigner transformation:
\begin{equation}
\begin{aligned}
\Gamma_1&=\sigma_x \otimes 1\otimes1\otimes1\otimes \cdots \otimes 1, &
\Gamma_2&=\sigma_y \otimes 1\otimes1\otimes1\otimes \cdots \otimes 1, \\
\Gamma_3&=\sigma_z \otimes \sigma_x \otimes 1\otimes1\otimes \cdots \otimes 1, &
\Gamma_4&=\sigma_z \otimes \sigma_y \otimes 1\otimes1\otimes \cdots \otimes 1, \\
\Gamma_5&=\sigma_z \otimes \sigma_z \otimes \sigma_x \otimes 1\otimes \cdots \otimes 1, &
\Gamma_6&=\sigma_z \otimes \sigma_z \otimes \sigma_y \otimes 1\otimes \cdots \otimes 1,\\
\vdots & & \vdots & 
\end{aligned}
\end{equation}
where $\sigma_{x,y,z}$ are the standard Pauli matrices.
Given an element $\vec v\in \bR^{24}$, 
we then introduce Feynman's slash notation: $\slashed v = v_i \Gamma_i$.

Now, any element $g\in SO(24)$
can be written as $\theta_i$ rotations
in 12 planes, each of which is spanned by a pair of orthonormal vectors
$v^{(i)}$ and $w^{(i)}$, $i=1,\ldots,12$.
We take the convention that $w^{(i)}$ 
is obtained from $v^{(i)}$ by a $+\pi/2$ rotation in the $i$-th plane.
Then, the action of $\pm\hat g\in Spin(24)$ on the spinor representation is
\begin{equation}
\prod_{i=1}^{12}
\left( \cos \frac{\theta_i}{2} +
\sin \frac{\theta_i}{2} \slashed{v}^{(i)} \slashed{w}^{(i)} \right).
\end{equation}
Note that $\theta_i$ is only defined modulo $2\pi$,
translating to the fact that $\hat g$ is only defined up to a sign.
We need a systematic method to fix this ambiguity.

\subsubsection{Resolving the sign ambiguity}

The sign ambiguity can be resolved as follows.
First, recall that the generators $g_1$ and $g_2$ of $\Co_0$ 
satisfy the  two relations listed in \eqref{conway-relations},
where the abbreviations $A=g_1$ and $B=g_2$ were used.
The first equation contains an odd number of $B$'s,
and the second equation contains an odd number of $A$'s.
Therefore, among the various lifts $(\pm \hat A,\pm \hat B)$ with four choices of signs in total,
only one choice of signs satisfies the two relations in $Spin(24)$.
This fixes the sign ambiguity of the lifts of $g_1$ and $g_2$
to $Spin(24)$, which we denote by $\hat g_1$ and $\hat g_2$.
Then the lift of $\Co_0$ within $Spin(24)$ is generated by $\hat g_1$ and $\hat g_2$.

As mentioned in the main text, 
there is a one-dimensional subspace of the spinor representation 
which is invariant under the action of $\Co_0$.
This subspace can be determined by finding the common fixed subspace
of $\hat g_1$ and $\hat g_2$,
and this is the subspace spanned by the supercurrent $Q$ 
of the Duncan theory.

Once the supercurrent $Q$ is determined,
the sign ambiguity of the lift of any $g\in \Co_0\subset SO(24)$
to $\hat g\in Spin(24)$ can be fixed by demanding that 
$+\hat g$ rather than $-\hat g$ leaves $Q$ invariant.
Given a commuting pair $(g,h)$ of $\Co_0\subset SO(24)$,
what we do is to simultaneously diagonalize $g$ and $h$
by finding 12 planes spanned by orthonormal vectors $v^{(i)}$ and $w^{(i)}$, $i=1,\ldots,12$,
such that $g$ and $h$ act as $\lambda_g^{(i)}$ and $\lambda_h^{(i)}$ rotations in these planes.
This allows us to construct $\pm \hat g$ and $\pm \hat h$
in $Spin(24)$.
We can then fix the sign ambiguity by demanding that $+\hat g$ and $+\hat h$ leave the supercurrent $Q$ invariant.

In this manner, we can fix $\lambda_{g,h}^{(1,\ldots,11)}$ modulo $2\pi$,
and $\lambda_{g,h}^{(12)}$ modulo $4\pi$.
In the actual computations, due to anomalies, 
we need to select concrete values for $\lambda_{g,h}^{(i)}$ as real numbers.

\subsection{Obtaining the partition functions}
\label{app:PF}
Once the computations up to this point are done,
we can implement the algorithm described in Sec.~\ref{sec:PFalgorithm}
to determine the anomaly-cancelling phases $A_\cO$ and $B_\cO$,
and then compute the partition functions of the orbifold theories
we are interested in.
This part of the computation is kept under the subdirectory \texttt{partition-function/} of the repository.
It is implemented in Mathematica.

The essence of the procedure is already given in Sec.~\ref{sec:PFalgorithm}.
Here we record some technical details of the implementation,
which were useful in doing the actual computations.
\begin{enumerate}
\item First, from each $SL(2,\bZ)$-orbit $\cO$ of commuting pairs of $\Co_0$,
we need to pick a representative commuting pair $(g_\cO,h_\cO)$.
It was useful to pick a representative $(g_\cO,h_\cO)$
such that, when $\langle g_\cO,h_\cO\rangle=\bZ_a \times \bZ_b$ with $a|b$,
$g_\cO$ is a generator of $\bZ_a$ and $h_\cO$ is a generator of $\bZ_b$.

\item Second, it was also useful to arrange representatives so that when $P \cO \neq \cO$,
we have \begin{equation}
(g_{P\cO},h_{P\cO})=(g_\cO,h_\cO^{-1}).
\end{equation}
It was also useful to arrange the lifted real-valued angles $(\vec\lambda,\vec\mu)$ to satisfy
\begin{equation}
\vec \lambda_{{P\cO}} = \vec\lambda_{\cO},\qquad
\vec \mu_{{P\cO}} = - \vec\mu_{\cO}.
\end{equation}

\item Third, it was useful to separate some `default values' from $\alpha(g_\cO,h_\cO)$ to define the anomaly-cancelling phases $A_\cO$ and $B_\cO$ in the following manner.
We assume that a representative commuting pair $(g_\cO,h_\cO)$ has been chosen
as described in the point 1 above.
Then, for orbits $\cO$ with $\cO=P\cO$,
we set \begin{equation}
\hat Z_g^h= A_\cO \eta_\cO e^{2\pi i 24 \vec \lambda\cdot \vec \mu} Z_{\vec\lambda}^{\vec\mu}.
\end{equation}
Here, the exponential factor is the phase \eqref{foofinal}
discussed in Sec.~\ref{sec:abelianDefault},
$\eta_\cO$ is an $a$-th root of unity allowed by the bosonic discrete torsion
so that $\eta_\cO e^{2\pi i 24 \vec \lambda\cdot \vec \mu} Z_{\vec\lambda}^{\vec\mu}$ is a real number,
and then $A_\cO=\pm1$ is the phase to be determined.
For orbits $\cO$ with $\cO\neq P\cO$,
we set \begin{equation}
  \hat Z_g^h= B_\cO  e^{2\pi i 24 \vec \lambda\cdot \vec \mu} Z_\lambda^\mu
\end{equation}
where $B_\cO$ is the phase to be determined.
With the choices of representatives $(g_\cO,h_\cO)$ and the lifted angles $\vec \lambda_{\cO}$ and $\vec \mu_{\cO}$ as described above in the point 2,
$B_\cO$ and $B_{P\cO}$ are complex conjugates of each other.
\end{enumerate}
With these choices, somewhat surprisingly,
most of $A_\cO$ and $B_\cO$ turned out to be simply $+1$.
Only 8 out of 639 $A_\cO$'s are $-1$,
and only five out of 30 pairs of $B_\cO$ and $B_{P\cO}$ are not $+1$.
The authors do not know any deeper significance of this fact,
but it definitely helped them while they were studying these phases
by trial and error.

\subsection{Investigating the mixed anomaly}
\label{app:mixed}

In this last subsection of the appendix, we describe 
how the group-theoretical and other computations
described in  Sec.~\ref{sec:mixed-A24} can be performed in practice.
This is kept under the subdirectory \texttt{mixed-anomaly/} of the repository,
and implemented in GAP and Mathematica. 
This subsection is mainly to illustrate how the actual computations are done,
to the readers who are not interested enough to actually clone the git repository
but are interested in getting the flavor of the techniques involved.

Suppose that permutation representations of the generators $g_1$ and $g_2$
for $\Co_1$, which should be computed as described in Appendix~\ref{app:Co}, 
are already saved in \texttt{co1-g1g2.txt}.
We construct $\Co_1$ inside GAP as follows: \begin{lstlisting}
Read("co1-g1g2.txt");
Co1:=Group(g1,g2);
\end{lstlisting}
Its conjugacy classes, their representatives,
their centralizers, and the conjugacy classes of the centralizers
can be routinely computed as follows:
\begin{lstlisting}
cc:=ConjugacyClasses(Co1);
conjrep:=List(cc,Representative);
centralizers:=[];
centralizerconjugacyclasses:=[];
for i in [2..101] do
	centralizers[i]:=Centralizer(Co1,conjrep[i]);
od;
centralizers[1]:=Co1;
for i in [2..101] do
	centralizerconjugacyclasses[i]:=
    ConjugacyClasses(centralizers[i]);
od;
centralizerconjugacyclasses[1]:=cc;
centralizerconjugacyclassreps:=	List(
  centralizerconjugacyclasses,x->List(x,Representative));
\end{lstlisting}
As the order of the conjugacy classes vary from run to run,
it is imperative to save the data in a file.
It is convenient to simply dump the workspace to a file as follows:
\begin{lstlisting}
SaveWorkspace("conway1");
quit;
\end{lstlisting}
For the subsequent computations, the workspace can be loaded using the \code{-L} option.

The mixed anomaly involves the commuting pair $\#(4,21)$ of $\Co_1$,
which we denote by $(g,h)$.
We want to know the conjugacy classes of commuting pairs $(g^{2x},h g^y)$
for $x=0,\ldots,12$ and $y=0,\ldots,25$.
This can be done by first defining the function to compute the conjugacy class of a given commuting pair:
\begin{lstlisting}
cr:=centralizerconjugacyclassreps;;
findclass := function(group,cc,elem)
	local i;
	for i in [1..Length(cc)] do
		if IsConjugate(group,cc[i],elem) then
			return i;
		fi;
	od;
end;

findComPair:=function(a,b)
	local i,c,d,j;
	i:=findclass(Co1,conjrep,a);
	c:=RepresentativeAction(Co1,conjrep[i],a);
	d:=c*b*Inverse(c);
	j:=findclass(centralizers[i],cr[i],d);
	return [i,j];
end;
\end{lstlisting}
We then simply loop over $x$ and $y$ to find the conjugacy classes of commuting pairs $(g^{2x},h g^y)$:
\begin{lstlisting}
g:=cr[1][4];
h:=cr[4][21];
result:=List([0..12],x->List([0..25],
  y->findComPair(g^(2*x),h*(g^y))));
PrintTo("result.txt",result);
\end{lstlisting}

We convert the output to the Mathematica format, save it to \texttt{result.m},
and investigate it using our precomputed data.
To do so, first load \texttt{partition-function-common.nb}
inside the directory \texttt{partition-function/} 
and run it.
This loads the projection which sends a $\Co_0$ commuting pair $\#(i,j)$
to the corresponding $\Co_1$ commuting pair $\#(i',j')$.
This information is kept in the variable \code{proj}.
We need to find the preimage instead,
so we define the function \code{lift} as follows:
\begin{lstlisting}[language=Mathematica]
Do[Do[
	lift[proj@[[i,j@]]]={i,j}, {j,1,Length[proj@[[i@]]]}
],{i,1,Length[proj]}];
\end{lstlisting}
Assuming that the content of \texttt{result.m} is read into a variable \code{X},
we would like to find  the preimages of the conjugacy classes of commuting pairs $(g^{2x},h g^y)$ of $\Co_1$ in $\Co_0$,
and then send them to their $SL(2,\bZ)$-orbit representatives.
This can be done as follows:
\begin{lstlisting}[language=Mathematica]
Flatten[Table[
   orbitRepresentative[lift[X@[[i+1, j+1@]]]], 
   {j,0,25}, {i,0,11}], 1] // Union
\end{lstlisting}
which returns \code{\{\{1,38\}, \{1,156\}\}}.
We can now load the precomputed data of the partition functions
of the Duncan theory on the torus with the holonomy of $\Co_0$
specified by a commuting pair $(g,h)$:
\begin{lstlisting}[language=Mathematica]
SinglePF[orbitRepresentative[{1,38}]]
SinglePF[orbitRepresentative[{1,156}]]
\end{lstlisting}
both of which return \code{0}.

This was the technical computation necessary to show the existence
of the mixed anomaly between $A_{24}$ and $\Co_1$ we discussed
in  Sec.~\ref{sec:mixed-A24}.
The file \texttt{partition-function-common.nb}
contains various other useful functions and data,
and the interested readers are encouraged to look into it.

 \bibliographystyle{ytamsalpha}
 \def\arxivfont{\rm}
 \baselineskip=.95\baselineskip
\bibliography{ref}

\end{document}